\documentclass[12pt,english]{article}
\usepackage{lmodern}
\usepackage[T1]{fontenc}
\usepackage[latin9]{inputenc}
\usepackage{geometry}
\usepackage{color}
\usepackage{babel}
\usepackage{mathrsfs}
\usepackage{mathtools}
\usepackage{amsmath}
\usepackage{amsthm}
\usepackage{amssymb}
\usepackage{graphicx}
\usepackage{setspace}
\usepackage[authoryear]{natbib}
\usepackage[unicode=true,
 bookmarks=true,bookmarksnumbered=false,bookmarksopen=false,
 breaklinks=false,pdfborder={0 0 0},pdfborderstyle={},backref=false,colorlinks=true]
 {hyperref}
\hypersetup{
 pdfpagelayout=OneColumn, pdfnewwindow=true, pdfstartview=XYZ, plainpages=false, urlcolor=[rgb]{0.0430 ,0, 0.5}, linkcolor=[rgb]{0.0430 ,0, 0.5}, citecolor=[rgb]{0.0430 ,0, 0.5}, hypertexnames=false}

\makeatletter
\providecommand{\tabularnewline}{\\}
\theoremstyle{definition}
\newtheorem{defn}{\protect\definitionname}
\theoremstyle{plain}
\newtheorem{prop}{\protect\propositionname}
\theoremstyle{definition}
 \newtheorem{example}{\protect\examplename}
\theoremstyle{plain}
\newtheorem{cor}{\protect\corollaryname}
\theoremstyle{plain}
\newtheorem{lem}{\protect\lemmaname}
\theoremstyle{plain}
\newtheorem{thm}{\protect\theoremname}

\usepackage{chngcntr}
\setcitestyle{round}
\usepackage{mathtools}

\makeatother

\providecommand{\corollaryname}{Corollary}
\providecommand{\definitionname}{Definition}
\providecommand{\examplename}{Example}
\providecommand{\lemmaname}{Lemma}
\providecommand{\propositionname}{Proposition}
\providecommand{\theoremname}{Theorem}

\begin{document}
\title{\vspace{-3.0cm}Payoff Information and Learning in Signaling Games\thanks{We thank Laura Doval, Glenn Ellison, Lorens Imhof, Yuichiro Kamada,
Robert Kleinberg, David K. Levine, Kevin K. Li, Eric Maskin, Dilip
Mookherjee, Harry Pei, Matthew Rabin, Bill Sandholm, Lones Smith,
Joel Sobel, Philipp Strack, Bruno Strulovici, Tomasz Strzalecki, Jean
Tirole, Juuso Toikka, and two anonymous referees for helpful comments
and conversations, and National Science Foundation grant SES 1643517
for financial support.}}
\author{Drew Fudenberg\thanks{Department of Economics, MIT. Email: \texttt{\protect\href{mailto:drew.fudenberg\%40gmail.com}{drew.fudenberg@gmail.com}}}
\and Kevin He\thanks{California Institute of Technology and University of Pennsylvania.
Email: \texttt{\protect\href{mailto:hesichao\%40gmail.com}{hesichao@gmail.com}}}}
\date{{\normalsize{}}%
\begin{tabular}{rl}
First version: & August 31, 2017\tabularnewline
This version: & November 25, 2019\tabularnewline
\end{tabular}}

\maketitle
\vspace{-0.27in}

\begin{abstract}
\begin{singlespace}
{\normalsize{}We add the assumption that players know their opponents'
payoff functions and rationality to a model of non-equilibrium learning
in signaling games. Agents are born into player roles and play against
random opponents every period. Inexperienced agents are uncertain
about the prevailing distribution of opponents' play, but believe
that opponents never choose conditionally dominated strategies. Agents
engage in active learning and update beliefs based on personal observations.
Payoff information can refine or expand learning predictions, since
patient young senders' experimentation incentives depend on which
receiver responses they deem plausible. We show that with payoff knowledge,
the limiting set of long-run learning outcomes is bounded above by
}\emph{\normalsize{}rationality-compatible equilibria}{\normalsize{}
(}\emph{\normalsize{}RCE}{\normalsize{}), and bounded below by }\emph{\normalsize{}uniform
RCE}{\normalsize{}. RCE refine the Intuitive Criterion \citep{cho_signaling_1987}
and include all divine equilibria \citep{banks_equilibrium_1987}.
Uniform RCE sometimes but not always exists, and implies universally
divine equilibrium.}{\normalsize\par}
\end{singlespace}

\bigskip{}


JEL classification codes C72, C73, D83

{\normalsize{}\thispagestyle{empty}
\setcounter{page}{0}}{\normalsize\par}
\end{abstract}
\begin{flushleft}
\interfootnotelinepenalty=10000
\renewcommand*\&{and}
\par\end{flushleft}

\section{Introduction}

Signaling games typically have many perfect Bayesian equilibria, because
Bayes rule does not pin down the receiver's off-path beliefs about
the sender's type. Different off-path beliefs for the receiver can
justify different off-path receiver behaviors, which in turn sustain
equilibria with a variety of on-path outcomes. For this reason, applied
work using signaling games typically invokes some equilibrium refinement
to obtain a smaller and (hopefully) more accurate subset of predictions.

However, most refinements impose restrictions on the off-path beliefs
without any reference to the process that might lead to equilibrium.
As in our earlier paper, \citet{fudenberg_he_2017}, this paper uses
a learning model to derive restrictions on out-of-equilibrium beliefs
and thus restrict the equilibrium set. The innovation here is to restrict
the agents' initial beliefs about opponents' strategies to reflect
knowledge of the opponents' utility functions (and that the opponents
act to maximize their expected utility). This generates restrictions
on long-run play that refine the Intuitive Criterion \citep{cho_signaling_1987}
and include all divine equilibria \citep{banks_equilibrium_1987}.

In our learning model, agents repeatedly play the same signaling game
against random opponents each period. Agents are Bayesians who believe
they face a fixed but unknown distribution of the opposing players'
strategies. Importantly, the senders start with independent prior
beliefs about how receivers respond to different signals, so they
cannot use the response to one signal to infer anything about the
distribution of responses to a different signal. This introduces an
exploration-exploitation trade-off, as each sender only observes the
response to the one signal she sends each period. Long-lived and patient
senders will therefore experiment with every signal that they think
might yield a substantially higher payoff than the myopically optimal
signal. Thus, ``out of equilibrium'' signals actually arise with
positive probability as experiments. The key to our results is that
different types of senders have different incentives for experimenting
with various signals, so that some of the sender types will send certain
signals more often than other types do. Consequently, even though
long-lived senders only experiment for a vanishingly small fraction
of their lifetimes, the play of the long-lived receivers will best
respond to beliefs about the senders' types that reflects this difference
in experimentation probabilities.

Of course, the senders' experimentation incentives depend on their
prior beliefs about which receiver responses are plausible after each
signal. The set of learning outcomes depends on the assumptions we
make on the set of ``allowable'' priors over opponents' strategies.
\citet{fudenberg_he_2017} restricted the set of allowable priors
to be \emph{non-doctrinaire, }which implies that agents always assign
a strictly positive probability to their opponents playing strictly
dominated strategies. An equilibrium profile is \emph{patiently stable
}if it is a long-run outcome with patient and long-lived agents for
some non-doctrinaire prior; \citet{fudenberg_he_2017} shows that
patiently stable profiles of signaling games must satisfy a condition
called the \emph{compatibility criterion} (CC).

In this paper, we instead assume that the players' prior beliefs encode
knowledge of their opponents' payoff functions, so in particular the
senders assign zero probability to the event that the receivers choose
conditionally dominated actions after any signal.\footnote{An action is \emph{conditionally dominated }after a given signal if
it does not best respond to any belief about the sender's type.} Inexperienced senders with full-support beliefs about the receivers'
play may experiment with a signal in the hope that the receivers respond
with a certain favorable action, not knowing that this action will
never be played as it is not a best response to any receiver belief.
With payoff information, even very patient senders will never undertake
such experiments. Conversely, receivers know that no sender type would
ever want to play a signal that does not best respond to any receiver
strategy, because no possible response by the receiver would make
playing that signal worthwhile. For this reason, the receivers' beliefs
after each signal assign probability zero to the types for whom that
signal is dominated.

We introduce equilibrium refinements for signaling games that provide
upper and lower bounds on the set of \emph{rationally patiently stable}
learning outcomes, the analog of patient stability when the allowable
priors reflect payoff knowledge and are otherwise non-doctrinaire.
Theorem \ref{thm:necessary} shows that every rationally patient learning
outcome where receivers have strict best responses to the on-path
signals must be a ``rationality compatible equilibrium'' (RCE);
this is an equilibrium where off-path beliefs satisfy restrictions
derived from the comparative experimentation frequencies of different
sender types when they know the receiver's payoff function. Conversely,
Theorem \ref{thm:sufficient} shows that every equilibrium that satisfies
a uniform version of rational compatibility (uRCE) and some strictness
assumptions can arise as a patient learning outcome. That is, for
equilibria satisfying the relevant strictness assumptions, we have
\[
\text{uRCE}\subseteq\text{rationally patiently stable profiles}\subseteq\text{\text{RCE}}
\]

Importantly, the set of allowable priors with payoff information are
not nested with the priors we considered in \citet{fudenberg_he_2017}:
Payoff information rules out priors that assign positive probability
to strictly dominated actions, while the full-support priors of \citet{fudenberg_he_2017}
require that strictly dominated strategies receive positive probability.
Thus the sets of patiently stable profiles and rationally patiently
stable profiles are not in general nested. Through a pair of examples,
we illustrate that RCE better tracks the set of rationally patiently
stable outcomes than the solution concepts from \citet{fudenberg_he_2017}
do.

In Example \ref{exa:charity}, the sender can be either strong or
weak. The sender has a safe option \textbf{Out} with a known payoff,
and a risky option \textbf{In} whose payoff depends on the receiver's
response. The receiver has three responses to \textbf{In}: \textbf{Up},
which is optimal against the strong sender; \textbf{Down}, which is
optimal against the weak sender, and \textbf{X}, which is never optimal.
We show that RCE refines \citet{fudenberg_he_2017}'s CC, and identifies
the unique rationally patiently stable strategy profile. Intuitively,
this is because when priors encode payoff information, the strong
types experiment more with \textbf{In} than the weak types do against
any play of the receivers, which is not true when senders do not know
the receivers' payoff functions.

In some other games, the set of rationally patiently stable profiles
is larger than that of patiently stable profiles, because payoff information
leads to some experiments not being taken at all. In Example \ref{exa:payoff_info_expands},
there is a signal that is never used as an experiment when senders
have payoff information, so the receivers' beliefs and behavior after
this signal are arbitrary. On the other hand, when senders are ignorant
of the receivers' payoff functions, one sender type will experiment
much more frequently with this signal than the other type, leading
to a restriction on the receivers' off-path beliefs and off-path behavior
after the signal. In this game, RCE exactly identifies the set of
rationally patiently stable strategy profiles, while the \emph{strong
CC} (which is shown by \citet{fudenberg_he_2017} to be necessary
for patient stability without payoff knowledge) rules out some rationally
patiently stable profiles.

Like RCE and uRCE, standard equilibrium refinements do reflect the
idea that players know that their opponents will not play strictly
dominated strategies. Moreover, as we explain in Section \ref{sec:Comparison-to-Other},
the nesting relationships
\[
\text{uRCE}\subsetneq\text{universally divine}\subsetneq\text{divine}\subsetneq\text{\text{RCE}}\subsetneq\text{Intuitive Criterion}\subsetneq\text{Nash}.
\]
hold for all equilibria where the receiver has strict incentives after
all on-path signals. In particular, our learning-based belief restrictions
resemble those imposed by divine equilibrium \citep{banks_equilibrium_1987}:
Every divine equilibrium is also an RCE and every uRCE is path-equivalent
to a universally divine equilibrium. We should point out, though,
that while every game has at least one rationally patiently stable
profile, a uRCE need not exist; Example \ref{exa:beer-quiche} is
one simple case where one does not.

\subsection{Related Literature}

This paper is most closely related to the work of \citet{fudenberg_steady_1993},
\citet{fudenberg_superstition_2006}, and \citet{fudenberg_he_2017}
on patient learning by Bayesian agents who believe they face a steady-state
distribution of play. Except for the support of the agents' priors,
our learning model is exactly the same as that of \citet{fudenberg_he_2017},
and the proof of Theorem \ref{thm:necessary} follows the lines of
our results there. Theorem \ref{thm:sufficient} is the main technical
innovation. It establishes a sufficient condition for an equilibrium
to be rationally patiently stable\emph{. }The proof of this sufficient
condition for rational patient stability constructs a suitable prior
and analyzes the corresponding rationally patiently stable profiles.
The only other constructive sufficient condition\footnote{``Constructive,'' as opposed to proofs that rule out all but one
equilibrium using necessary conditions and then appeal to an existence
theorem for patiently stable steady states. Constructive sufficient
conditions allow us to characterize learning outcomes more precisely
in games where multiple equilibria satisfy the necessary conditions,
such as Example \ref{exa:education}.} for strategy profiles to be patiently stable is Theorem 5.5 of \citet{fudenberg_superstition_2006},
which only applies to a subclass of perfect-information games. In
such games the relative probabilities of various off-path actions
do not matter, because each off-path experiment is perfectly revealed
when it occurs. Indeed, the central lemma leading to Theorem \ref{thm:sufficient}
constructs a prior belief to ensure that the receivers correctly learn
the relative frequencies that different types undertake various off-path
experiments. This lemma deals with an issue specific to signaling
games, and is not implied by any result in \citet{fudenberg_superstition_2006}.

Like this paper, \citet{cho_signaling_1987} and \citet{banks_equilibrium_1987}
study equilibrium refinements in signaling games. We compare our learning-based
equilibrium refinements with their refinements, which implicitly assume
that players are certain of the payoff functions of their opponents.

Our paper is also related to other models of Bayesian non-equilibrium
learning, such as \citet{kalai_rational_1993} and \citet{esponda_berknash_2016},
though these papers do not study optimal experimentation and do not
refine the Nash equilibrium set. Finally, \citet*{dekel1999payoff},
\citet{fudenberg_kamada_2015_rationalizable}, and \citet{fudenberg2018rationalizable}
develop equilibrium refinements that combine the idea of equilibrium
arising from learning with the assumption that players know one another's
payoff functions and feedback structures. These paper do no provide
explicit learning models, but the implicit models they have in mind
would feature impatient learners who do little or no experimentation.

\section{\label{sec:two_refinements}Two Equilibrium Refinements for Signaling
Games}

\subsection{Signaling Game Notation}

A \emph{signaling game} has two players, a sender (``she,'' player
1) and a receiver (``he,'' player 2). At the start of the game,
the sender learns her type $\theta\in\Theta$, but the receiver only
knows the sender's type distribution\footnote{The notation $\Delta(X)$ means the set of all probability distributions
on $X$.} $\lambda\in\Delta(\Theta)$. Next, the sender chooses a signal $s\in S$.
The receiver observes $s$ and chooses an action $a\in A$ in response.
We assume that $\Theta,S,A$ are finite and that $\lambda(\theta)>0$
for all $\theta.$

The players' payoffs depend on the triple $(\theta,s,a)$. Let $u_{1}:\Theta\times S\times A\to\mathbb{R}$
and $u_{2}:\Theta\times S\times A\to\mathbb{R}$ denote the utility
functions of the sender and the receiver, respectively.

For $P\subseteq\Delta(\Theta)$, we have
\[
\text{BR}(P,s)\coloneqq\bigcup_{p\in P}\left(\underset{a\in A}{\arg\max}\ \mathbb{E}_{\theta\sim p}\left[\text{\ensuremath{u_{2}(\theta,s,a)}}\right]\right)
\]
 as the set of best responses to $s$ supported by some belief in
$P$. Letting $P=\Delta(\Theta)$, the set $A_{s}^{\text{BR}}:=\text{BR}(\Delta(\Theta),s)\subseteq A$
contains the receiver actions that best respond to some belief about
the sender's type after $s$. We say that actions in $A_{s}^{\text{BR}}$
are \emph{conditionally undominated after signal $s$,} and that actions
in $A\backslash A_{s}^{\text{BR}}$ are \emph{conditionally dominated
after signal $s$.} We denote by $\Pi_{2}^{\bullet}\coloneqq\times_{s\in S}\Delta(A_{s}^{\text{BR}})$
the \emph{rational receiver strategies; }these are the strategies
that assign probability 0 to conditionally dominated actions.\emph{}\footnote{Throughout we adopt the terminology ``strategies'' to mean behavior
strategies, not mixed strategies.}\emph{ }The rational receiver strategies form a subset of $\Pi_{2}:=\times_{s\in S}\Delta(A)$,
the set of all receiver strategies. A sender who knows the receiver's
payoff function expects the receiver to choose a strategy in $\Pi_{2}^{\bullet}$.

A\emph{ sender strategy} $\pi_{1}=$ $(\pi_{1}(\cdot\mid\theta))_{\theta\in\Theta}\in\Pi_{1}$
specifies a distribution on $S$ for each type, $\pi_{1}(\cdot\mid\theta)\in\Delta(S)$$.$
For a given $\pi_{1}$, signal $s$ is` \emph{off the path of play
}if it has probability 0, i.e. $\pi_{1}(s\mid\theta)=0$ for all $\theta.$
Let 
\[
S_{\theta}:=\bigcup_{\pi_{2}\in\Pi_{2}}\left(\underset{s\in S}{\arg\max}\ u_{1}(\theta,s,\pi_{2}(\cdot\mid s))\right).
\]
be the set of signals that best respond to some (not necessarily rational)
receiver strategy for type $\theta$. Signals in $S\backslash S_{\theta}$
are \emph{dominated} for type $\theta$, and $\Pi_{1}^{\bullet}\coloneqq\times_{\theta}\Delta\left(S_{\theta}\right)$
denotes the \emph{rational sender strategies} where no type ever sends
a dominated signal. We also write $\Theta_{s}$ for the types $\theta$
for whom $s\in S$ is not dominated. A receiver who knows the sender's
payoff function expects the sender to choose a strategy in $\Pi_{1}^{\bullet}$
and only expects types in $\Theta_{s}$ to play signal $s$.

\subsection{Rationality-Compatible Equilibria}

We now introduce rationality-compatible equilibrium (RCE) and uniform
rationality-compatible equilibrium (uRCE), two refinements of Nash
equilibrium in signaling games.

In Section \ref{sec:learning_model}, we develop a steady-state learning
model where populations of senders and receivers, initially uncertain
as to the aggregate play of the opponent population, undergo random
anonymous matching each period to play the signaling game. We study
the steady states when agents are patient and long lived, which we
term \emph{rationally patiently stable}. Under some strictness assumptions,
we show that only RCE can be rationally patiently stable (Theorem
\ref{thm:necessary}) and that every uRCE is ``path-equivalent''\footnote{Roughly speaking, this means equivalent up to changing some of the
receiver's off-path behavior, see Subsection \ref{subsec:Divine-Equilibrium}.} to a rationally patiently stable profile (Theorem \ref{thm:sufficient}).
Thus we provide a learning foundation for these solution concepts.

Our learning foundation will assume that agents know other agents'
utility functions and know that other agents are rational in the sense
of playing strategies that maximize the corresponding expected utilities.
We will not however iteratively assume higher orders of payoff knowledge
and rationality, so that we model ``rationality'' as opposed to
``rationalizability.''\footnote{It is straightforward to extend our results about RCE to priors that
reflect higher-order knowledge of the rationality and payoff functions
of the other player. The resulting equilibrium refinement always exists,
and like RCE is implied by universal divinity. We do not include it
here both because we are unaware of any interesting examples where
the additional power has bite, and because we are skeptical about
the hypothesis of iterated rationality.}

In the learning model, this implies senders' uncertainty about receivers'
play is always supported on $\Pi_{2}^{\bullet}$ instead of $\Pi_{2}$,
and similarly receivers' uncertainty about senders' play is supported
on $\Pi_{1}^{\bullet}$ instead of $\Pi_{1}$. In Section \ref{subsec:knowing_utilities_examples},
we discuss heuristically how our solution concepts capture some of
the ways in which payoff information affects learning outcomes. This
discussion will later be formalized in the context of the learning
model we develop in Section \ref{sec:learning_model}.
\begin{defn}
\label{def:compatible_order} Signal $s$ is \emph{more rationally-compatible}
with $\theta^{'}$ than $\theta^{''}$, written as $\theta^{'}\succsim_{s}\theta^{''}$,
if for every $\pi_{2}\in\Pi_{2}^{\bullet}$ such that 
\[
u_{1}(\theta^{''},s,\pi_{2}(\cdot|s))\ge\max_{s^{'}\ne s}u_{1}(\theta^{''},s^{'},\pi_{2}(\cdot|s^{'})),
\]
we have 
\[
u_{1}(\theta^{'},s,\pi_{2}(\cdot|s))>\max_{s^{'}\ne s}u_{1}(\theta^{'},s^{'},\pi_{2}(\cdot|s^{'})).
\]
\end{defn}
In words, $\theta^{'}\succsim_{s}\theta^{''}$ means whenever $s$
is a weak best response for $\theta^{''}$ against some rational receiver
behavior strategy $\pi_{2}$, it is a strict best response for $\theta^{'}$
against $\pi_{2}$.

The next proposition shows that $\succsim_{s}$ is transitive and
``almost'' asymmetric. A signal $s$ is \emph{rationally strictly
dominant }for $\theta$ if it is a strict best response against any
rational receiver strategy, $\pi_{2}\in\Pi_{2}^{\bullet}$. A signal
$s$ is \emph{rationally strictly dominated }for $\theta$ if it is
not a weak best response against any rational receiver strategy.
\begin{prop}
\label{prop:transitive_asymm} We have
\end{prop}
\begin{enumerate}
\item $\succsim_{s^{'}}$ is transitive.
\item Except when $s^{'}$ is either rationally strictly dominant for both
$\theta^{'}$ and $\theta^{''}$ or rationally strictly dominated
for both $\theta^{'}$ and $\theta^{''}$, $\theta^{'}\succsim_{s^{'}}\theta^{''}$
implies $\theta^{''}\not\succsim_{s^{'}}\theta^{'}$.
\end{enumerate}
The Appendix provides proofs for all of our results except where otherwise
noted.

We require two auxiliary definitions before defining RCE.
\begin{defn}
\label{def:odds_ratio_set} For any two types $\theta^{'},\theta^{''}$,
let $P_{\theta^{'}\triangleright\theta^{''}}$ be the set of beliefs
where the odds ratio of $\theta^{'}$ to $\theta^{''}$ exceeds their
prior odds ratio, that is\footnote{With the convention $\frac{0}{0}:=0$.}
\begin{equation}
P_{\theta^{'}\triangleright\theta^{''}}:=\left\{ p\in\Delta(\Theta):\frac{p(\theta^{''})}{p(\theta^{'})}\le\frac{\lambda(\theta^{''})}{\lambda(\theta^{'})}\right\} .\label{eq:odds_ratio_P}
\end{equation}
\end{defn}
Note that if $\pi_{1}(s|\theta^{'})\ge\pi_{1}(s|\theta^{''}),$ $\pi_{1}(s|\theta^{'})>0,$
and the receiver updates beliefs using $\pi_{1}$, then the receiver's
posterior belief about the sender's type after observing $s$ falls
in the set $P_{\theta^{'}\triangleright\theta^{''}}$. In particular,
in any Bayesian Nash equilibrium, the receiver's on-path belief falls
in $P_{\theta^{'}\triangleright\theta^{''}}$ after any on-path signal
$s$ with $\theta^{'}\succsim_{s}\theta^{''}$.

We now introduce some additional definitions to let us investigate
the implications of the agents' knowledge of their opponent's payoff
function. For a strategy profile $\pi^{*}$, let $\mathbb{E}_{\pi^{*}}[u_{1}\mid\theta]$
denote type $\theta$'s expected payoff under $\pi^{*}$.
\begin{defn}
\label{def:J-tilde}For any strategy profile $\pi^{*}$, let
\[
\widetilde{J}(s,\pi^{*})\coloneqq\left\{ \theta\in\Theta:\underset{a\in A_{s}^{\text{BR}}}{\max}\ u_{1}(\theta,s,a)\geq\mathbb{E}_{\pi^{*}}[u_{1}\mid\theta]\right\} .
\]
\end{defn}
This is the set of types for which \emph{some }best response to signal
$s$ is at least as good as their payoff under $\pi^{*}.$ For all
other types, the signal $s$ is equilibrium dominated in the sense
of \citet{cho_signaling_1987}.
\begin{defn}
\label{def:RCE-beliefs} The set of\emph{ rationality-compatible beliefs
}for the receiver at strategy profile $\pi^{*}$, $\left(\tilde{P}(s,\pi^{*})\right)_{s},$
is defined as follows:

\begin{align*}
\begin{cases}
\tilde{P}(s,\pi^{*})\coloneqq\Delta(\widetilde{J}(s,\pi^{*}))\bigcap\left(\bigcap\limits _{(\theta^{'},\theta^{''})\text{ s.t. }\theta^{'}\succsim_{s}\theta^{''}}P_{\theta^{'}\triangleright\theta^{''}}\right) & \text{if }\widetilde{J}(s,\pi^{*})\ne\varnothing\\
\tilde{P}(s,\pi^{*})\coloneqq\Delta(\Theta_{s}) & \text{if }\widetilde{J}(s,\pi^{*})=\varnothing.
\end{cases}
\end{align*}

\end{defn}
The main idea behind the rationality-compatible beliefs is that the
receiver's posterior likelihood ratio for types $\theta^{'}$ and
$\theta^{''}$ dominates the prior likelihood ratio whenever $\theta^{'}\succsim_{s}\theta^{''}$.
A second feature involves equilibrium dominance. Note that $\widetilde{P}$
assigns probability 0 to equilibrium-dominated types; this is similar
to the belief restriction of the Intuitive Criterion. Note that this
definition imposes no belief restrictions based on $\theta^{'}\succsim_{s}\theta^{''}$
when $s$ is equilibrium dominated for every type. As we illustrate
in Example \ref{exa:payoff_info_expands}, the receiver needs not
learn the rational compatibility relation when equilibrium dominance
leads to steady states where no type ever experiments with a certain
signal.
\begin{defn}
Strategy profile $\pi^{*}$ is a\emph{ rationality-compatible equilibrium
(RCE) }if it is a Nash equilibrium and $\pi_{2}^{*}(\cdot\mid s)\in\Delta(\text{BR}(\tilde{P}(s,\pi^{*}),s))$
for every $s$.
\end{defn}
RCE requires that the receiver only plays best responses to rationality-compatible
beliefs after each signal. This solution concept allows for the possibility
that after off-path signals the receiver's strategy $\pi_{2}^{*}(\cdot\mid s)$
may not correspond to a single belief about the sender's type. We
show below that rationally patiently stable profiles exist and that
every rationally patiently stable profile is a RCE, so that RCE exist
as well.

Theorem \ref{thm:necessary} shows that RCE is a necessary condition
for a strategy profile where receivers have strict preferences after
each on-path signal to be rationally patiently stable. Intuitively,
this result holds because the optimal experimentation behavior of
the senders respects the compatibility order, and because, since players
eventually learn the equilibrium path, types will not experiment much
with signals that are equilibrium dominated. As we show in Section
\ref{sec:Comparison-to-Other}, RCE rules out the implausible equilibria
in a number of games, but is weaker than some past signaling game
refinements in the literature. However, RCE is only a necessary condition
for rational patient stability, which leaves open the question of
whether patient learning has additional implications. For this reason,
we now define uRCE, a subset of RCE (up to path-equivalence). As we
show below, uRCE is a sufficient condition for rational patient stability.
\begin{defn}
The set of\emph{ uniformly rationality-compatible beliefs }for the
receiver is $\left(\hat{P}(s)\right)_{s}$ where 
\[
\hat{P}(s)\coloneqq\Delta(\Theta_{s})\bigcap\left(\bigcap_{(\theta^{'},\theta^{''})\text{ s.t. }\theta^{'}\succsim_{s}\theta^{''}}P_{\theta^{'}\triangleright\theta^{''}}\right).
\]
\end{defn}
Note that $\left(\hat{P}(s)\right)_{s}$ makes no reference to a particular
strategy profile, unlike $\left(\tilde{P}(s,\pi^{*})\right)_{s}$.
Since $\Delta(\Theta_{s})$ contains types for whom $s$ is undominated
and $\widetilde{J}(s,\pi^{*})$ contains types for whom $s$ is equilibrium-undominated
(relative to the profile $\pi^{*}),$ we have $\tilde{P}(s,\pi^{*})\subseteq\hat{P}(s)$
whenever $\widetilde{J}(s,\pi^{*})\ne\varnothing$.
\begin{defn}
A Nash equilibrium strategy profile $\pi^{*}$ is called a \emph{uniform
rationality-compatible equilibrium} \emph{(uRCE)} if for all $\theta,$
all off-path signals $s$ and all $a\in\text{BR}(\hat{P}(s),s)$,
we have $\mathbb{E}_{\pi^{*}}[u_{1}\mid\theta]\ge u_{1}(\theta,s,a)$.

The ``uniformity'' in uniform RCE comes from the requirement that
\emph{every }best response to \emph{every} belief in $\hat{P}(s)$
deters \emph{every} type from deviating to the off-path $s$. By contrast,
a RCE is a Nash equilibrium where \emph{some} best response to $\tilde{P}(s,\pi^{*})$
deters every type from deviating to $s$. Unlike with RCE, uRCE need
not exist (see Example \ref{exa:beer-quiche}).

As the names imply, uRCE is a stronger solution concept than RCE,
up to path-equivalence.
\end{defn}
\begin{prop}
\label{prop:uRCE_RCE_path}Every uRCE is path-equivalent to an RCE.
\end{prop}

\subsection{\label{subsec:knowing_utilities_examples} Examples}

In this subsection, we show how to apply RCE and uRCE in specific
games, and compare them to the solution concepts from \citet{fudenberg_he_2017},
which are based on necessary conditions for patient stability with
full-support priors.

The following example illustrates that uRCE is a strict subset of
RCE in some games.
\begin{example}
\noindent \label{exa:education} Suppose a worker has either high
ability $(\theta_{H})$ or low ability $(\theta_{L})$. She chooses
between three levels of higher education: \textbf{None} (\textbf{N}),
\textbf{College} (\textbf{C}), or \textbf{Ph.D.} (\textbf{D}). An
employer observes the worker's education level and pays a wage, $a\in\{\text{\textbf{low}},\text{ \textbf{med}, \text{\textbf{high}}}\}$.
The worker's utility function is separable between wage and (ability,
education) pair, with $u_{1}(\theta,s,a)=z(a)+v(\theta,s)$ where
$z(\text{{\bf low}})=0,$ $z(\text{{\bf med}})=6,$ $z(\text{{\bf high}})=9$
and $v(\theta_{H},\text{{\bf N}})=0$, $v(\theta_{L},\boldsymbol{{\bf N}})=0$,
$v(\theta_{H},\text{{\bf C}})=2$, $v(\theta_{L},\text{{\bf C}})=1$,
$v(\theta_{H},\text{{\bf D}})=-2$, $v(\theta_{L},\text{{\bf D}})=-4$.
(With this payoff function, going to college has a consumption value
while getting a Ph.D. is costly.) The employer's payoffs reflect a
desire to pay a wage corresponding to the worker's ability and increased
productivity with education, given in the tables below.

\noindent \begin{center}

\noindent \medskip{}
\begin{tabular}{|c|c|c|c|}
\hline 
\textbf{N} & \textbf{low} & \textbf{med} & \textbf{high}\tabularnewline
\hline 
\hline 
$\theta_{H}$ & 0,-2 & 6,0 & 9,1\tabularnewline
\hline 
$\theta_{L}$ & 0,1 & 6,0 & 9,-2\tabularnewline
\hline 
\end{tabular}$\quad$%
\begin{tabular}{|c|c|c|c|}
\hline 
\textbf{C} & \textbf{low} & \textbf{med} & \textbf{high}\tabularnewline
\hline 
\hline 
$\theta_{H}$ & 2,-1 & 8,1 & 11,2\tabularnewline
\hline 
$\theta_{L}$ & 1,2 & 7,1 & 10,-1\tabularnewline
\hline 
\end{tabular}$\quad$%
\begin{tabular}{|c|c|c|c|}
\hline 
\textbf{D} & \textbf{low} & \textbf{med} & \textbf{high}\tabularnewline
\hline 
\hline 
$\theta_{H}$ & -2,0 & 4,2 & 7,3\tabularnewline
\hline 
$\theta_{L}$ & -4,3 & 2,2 & 5,0\tabularnewline
\hline 
\end{tabular}

\noindent \medskip{}
\end{center}

No education level is dominated for either type and no wage is conditionally
dominated after any signal. Since $v(\theta_{H},\cdot)-v(\theta_{L},\cdot)$
is maximized at \textbf{D}, it is simple to verify that $\theta_{H}\succsim_{\text{{\bf D}}}\theta_{L}$.
Similarly, $\theta_{L}\succsim_{\text{{\bf N}}}\theta_{H}$. There
is no compatibility relation at signal \textbf{C}.

When the prior is $\lambda(\theta_{H})=0.5$, the strategy profile
where the employer always pays a medium wage and both types of worker
choose \textbf{C} is a uRCE. This is because $\hat{P}(\text{{\bf N}})$
contains only those beliefs with $p(\theta_{H})\le0.5$, so $\text{BR}(\hat{P}(\text{{\bf N}}),\text{{\bf N}})=\{\text{\textbf{low}, \textbf{med}}\}$.
Both of these wages deter every type from deviating to \textbf{N}.
At the same time, no type wants to deviate to \textbf{D}, even if
she gets paid the best wage.

On the other hand, the equilibrium $\pi^{*}$ where the employer pays
low wages for \textbf{N} and \textbf{C}, a medium wage for \textbf{D},
and both types choose \textbf{D} is an RCE but not a uRCE. The belief
that puts probability 1 on the worker being $\theta_{L}$ belongs
to $\tilde{P}(\text{{\bf N}},\pi^{*})$ and $\tilde{P}(\text{{\bf C}},\pi^{*})$
and induces the employer to choose low wage. However, medium salary
is a best response to $\lambda\in\hat{P}(\text{{\bf N}})$ and medium
wage would tempt type $\theta_{L}$ to deviate to \textbf{N}. \hfill{}
$\blacklozenge$
\end{example}
In the learning model of \citet{fudenberg_he_2017}, agents do not
know others' utility functions and have full-support prior beliefs
about others' play. That paper's compatibility criterion (CC) is based
on a family of binary relations on types (one for each signal $s$)
that are less complete than the rational compatibility relations,
because the CC requires the condition that ``whenever $s$ is a weak
best response for $\theta^{''}$, it is also a strict best response
for $\theta^{'}$'' for all $\pi_{2}\in\Pi_{2}$ instead of only
for $\pi_{2}\in\Pi_{2}^{\bullet}$. Hence, RCE is always at least
as restrictive as the CC, and RCE can eliminate some equilibria that
the CC allows.
\begin{example}
\label{exa:charity} Consider a game where the sender has type distribution
$\lambda(\theta_{\text{strong}})=0.9,$ $\lambda(\theta_{\text{weak}})=0.1$
and chooses between two signals \textbf{In} or \textbf{Out}. The game
ends with payoffs (0,0) if the sender chooses \textbf{Out}. If the
sender chooses \textbf{In,} the receiver then chooses \textbf{Up},
\textbf{Down}, or \textbf{X}. \textbf{Up} is the receiver's optimal
response if the sender is more likely to be $\theta_{\text{strong}}$,
\textbf{Down} is optimal when the sender is more likely to be $\theta_{\text{weak}}$,
and \textbf{X} is never optimal.\footnote{This is a modified version of \citet{cho_signaling_1987}'s ``beer-quiche
game,'' where an outside option with certain payoffs (\textbf{Out})
replaces the \textbf{Quiche} signal. The responses \textbf{Up} and
\textbf{Down} correspond to \textbf{Not Fight} and \textbf{Fight}
in the beer-quiche game, while \textbf{X }is a conditionally dominated
response for the receiver following \textbf{In}. Also, while our definition
of signaling games requires that the receiver has the same action
set after every signal, this situation is clearly equivalent to one
where the receiver chooses \textbf{Up}, \textbf{Down}, or \textbf{X}
after \textbf{Out}, but all of these choices lead to the payoffs (0,0).} This game has two sequential equilibrium outcomes: one involving
both types choosing \textbf{Out}, and another where both types go
\textbf{In} and the receiver responds with \textbf{Up}.

\begin{center}\includegraphics[scale=0.3]{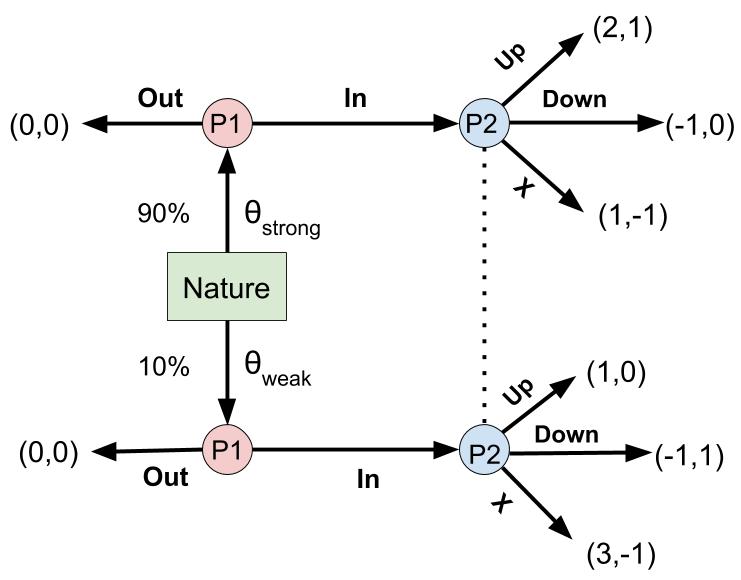}\end{center}

The sequential equilibrium outcome \textbf{Out} satisfies the CC,
because the compatibility relation of \citet{fudenberg_he_2017} does
not rank the two types after signal \textbf{In}. (For example, if
$\pi_{2}(\textbf{Down}\mid\textbf{In})=2/3$ and $\pi_{2}(\textbf{X}\mid\textbf{In})=1/3$,
$\theta_{\text{weak}}$ finds \textbf{In} optimal but $\theta_{\text{strong}}$
does not.)

However, since the stronger rational compatibility relation ranks
$\theta_{\text{strong}}\succsim_{\text{{\bf In}}}\theta_{\text{weak}}$,
the unique RCE is the equilibrium where both types go \textbf{In},
which implies that this is also the unique rationally patiently stable
outcome.\footnote{RCE requires that that $p(\theta_{\text{strong}}\mid\boldsymbol{\text{In}})\ge\lambda(\theta_{\text{strong}})=0.9,$
which implies that $\pi_{2}(\textbf{Up}\mid\textbf{In})=1$ in every
RCE. Therefore both types must be playing \textbf{In} in RCE.} \hfill{} $\blacklozenge$
\end{example}
The previous example shows how RCE can exclude some equilibria that
satisfy the CC. The next one cautions that RCE may allow more equilibrium
profiles than the strong compatibility criterion (strong CC), which
\citet{fudenberg_he_2017} show to be another necessary condition
for patient stability (with full-support priors). The strong CC requires
the receiver to put zero probability after signal $s$ on sender types
for whom $s$ is equilibrium dominated, but unlike in Definition \ref{def:J-tilde},
in the strong CC equilibrium dominance is computed by comparing the
type's equilibrium payoff to that of any response to the unsent signal,
including responses that are conditionally dominated.
\begin{example}
\label{exa:payoff_info_expands} Consider a game with two sender types,
$\theta_{1}$ and $\theta_{2}$, equally likely, and two possible
signals, \textbf{L} or \textbf{R.} Payoffs are given in the tables
below.

\begin{center}%
\begin{tabular}{|c|c|c|c|}
\hline 
signal: \textbf{L} & action: $a_{1}$ & action: $a_{2}$ & action: $a_{3}$\tabularnewline
\hline 
\hline 
type: $\theta_{1}$ & $-2$, 0 & $2,$ 2 & $2,$ 1\tabularnewline
\hline 
type: $\theta_{2}$ & $-2$, 1 & 2, 0 & 2, -1\tabularnewline
\hline 
\end{tabular}

\medskip{}

\begin{tabular}{|c|c|c|c|}
\hline 
signal: \textbf{R} & action: $a_{1}$ & action: $a_{2}$ & action: $a_{3}$\tabularnewline
\hline 
\hline 
type: $\theta_{1}$ & 5, -1 & -3, 2 & -4, 0\tabularnewline
\hline 
type: $\theta_{2}$ & -2, -1 & 1, 0 & 0, 1\tabularnewline
\hline 
\end{tabular}\end{center}

Action $a_{1}$ is conditionally dominated for the receiver after
signal \textbf{R}. It is easy to see that in every perfect Bayesian
equilibrium $\pi^{*}$, we must have $\pi_{1}^{*}(\text{{\bf L}}\mid\theta_{1})=\pi_{1}^{*}({\bf L}\mid\theta_{2})=1,$
$\pi_{2}^{*}(a_{2}\mid{\bf L})=1,$ and that $\pi_{2}^{*}(\cdot\mid\text{{\bf R}})$
must be supported on $A_{{\bf R}}^{\text{BR}}=\{a_{2},a_{3}\}$. This
means the off-path signal \textbf{R} is equilibrium dominated for
every type in $\pi^{*}$ (when they know the receiver's payoffs),
i.e. $\tilde{J}({\bf R},\pi^{*})=\varnothing.$ So, $\tilde{P}({\bf R},\pi^{*})=\Delta(\Theta_{\text{{\bf R}}})=\Delta(\Theta)$
and RCE permits the receiver to play either $a_{2}$ or $a_{3}$ after
\textbf{R}. (This is despite the fact that $\theta_{2}$ is more rationally
compatible with \textbf{R} than $\theta_{1}$ is. As we discussed
after Definition \ref{def:RCE-beliefs}, RCE does not restrict the
receiver's belief based on rational type compatibility after an off-path
signal that is equilibrium dominated for every type.)

By contrast, the strong CC from \citet{fudenberg_he_2017} requires
that the receiver plays $a_{2}$ after \textbf{R}: the equilibrium
payoff of both types is 2. Type $\theta_{1}$ has $\max_{a\in A}u_{1}(\theta_{1},\boldsymbol{\text{R}},a)\ge2$
but this is not true for $\theta_{2}.$ So the strong CC requires
the receiver to put probability 1 on $\theta_{1}$ after \textbf{R},
which pins down the receiver's off-path play.

We will show in Section \ref{subsec:Payoff-Information-and-Steady-State}
that when learners have payoff information, there is a rationally
patiently stable profile where the receivers play $a_{2}$ after \textbf{R}
and another rationally patiently stable profile where the receivers
respond to \textbf{R} with $a_{3}$. However, we will also show that
without payoff information, patient stability requires that the receivers
play $a_{2}$ after \textbf{R}. \hfill{} $\blacklozenge$
\end{example}
Finally, we show that uRCE may not exist.
\begin{example}
\noindent \label{exa:beer-quiche}\begin{center}

\includegraphics[scale=0.25]{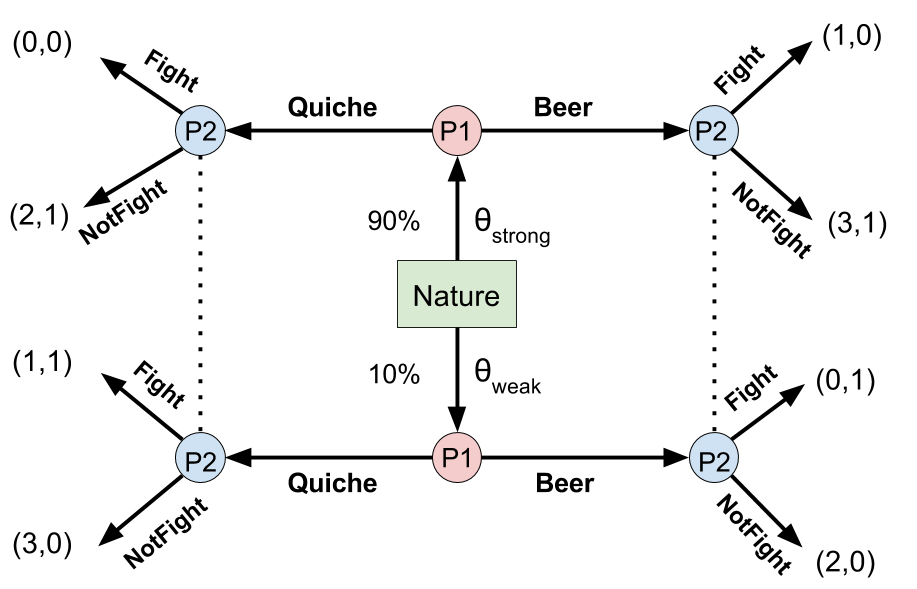}

\noindent \end{center}

In \citet{cho_signaling_1987}'s ``beer-quiche game,'' it is easy
to verify that the pooling equilibrium on \textbf{Beer} is the unique
RCE. This equilibrium is not a uRCE because $\hat{P}(\text{\textbf{Quiche}})=\{p:p(\theta_{\text{weak}})\ge0.1\}$,
so \textbf{NotFight} is a best response to a belief in $\hat{P}(\text{\textbf{Quiche}})$
that does not deter $\theta_{\text{weak}}$ from deviating. So the
game does not have any uRCE. Moreover, it is easy to see that the
same conclusions hold with slightly different assignments of payoffs
to terminal nodes, so that the non-existence result applies to an
open set of games.\hfill{} $\blacklozenge$
\end{example}

\section{\label{sec:Comparison-to-Other}Comparison to Other Equilibrium Refinements}

This section compares RCE to other equilibrium refinement concepts
in the literature.

\subsection{Iterated dominance}

We first relate RCE to a form of iterated dominance in the ex-ante
strategic form of the game, where the sender chooses a signal $\pi_{1}$
as function of her type. We show that every sender strategy that specifies
playing signal $s$ as a less compatible type $\theta^{''}$ but not
as a more compatible type $\theta^{'}$ will be removed by iterated
deletion. The idea is that such a strategy is never a weak best response
to any receiver strategy in $\Pi_{2}^{\bullet}$: if the less compatible
$\theta^{''}$ does not have a profitable deviation, then the more
compatible type strictly prefers deviating to $s$.
\begin{prop}
\label{prop:compatible_and_dominance} Suppose $\theta^{'}\succsim_{s}\theta^{''}$.
Then any ex-ante strategy of the sender $\pi_{1}$ with $\pi_{1}(s|\theta^{''})>0$
but $\pi_{1}(s|\theta^{'})<1$ is removed by strict dominance once
the receiver is restricted to using strategies in $\Pi_{2}^{\bullet}$.
\end{prop}

\subsection{The Intuitive Criterion}

We next relate RCE to the Intuitive Criterion.
\begin{prop}
\label{prop:IC}Every RCE satisfies the Intuitive Criterion.
\end{prop}
The next example shows that the set of RCE is strictly smaller than
the set of equilibria that pass the Intuitive Criterion.\footnote{\citet{fudenberg_he_2017}'s compatibility criterion is not always
more stringent than the Intuitive Criterion, but the strong compatibility
criterion studied in the same paper is.} The idea is that the Intuitive Criterion does not impose any restriction
on the relative likelihood of two types after a signal that is not
equilibrium dominated for either of them, but RCE can.
\begin{example}
\label{exa:IC_not_RCE} Consider a signaling game where the prior
probabilities of the two types are $\lambda(\theta^{'})=3/4$ and
$\lambda(\theta^{''})=1/4$, and the payoffs are:

\noindent \begin{center}\medskip{}
\begin{tabular}{|c|c|c|}
\hline 
signal: $s^{'}$ & action: $a^{'}$ & action: $a^{''}$\tabularnewline
\hline 
\hline 
type: $\theta^{'}$ & 4, 1 & 0, 0\tabularnewline
\hline 
type: $\theta^{''}$ & 6, 0 & 2, 1\tabularnewline
\hline 
\end{tabular} $\quad$%
\begin{tabular}{|c|c|c|}
\hline 
signal: $s^{''}$ & action: $a^{'}$ & action: $a^{''}$\tabularnewline
\hline 
\hline 
type: $\theta^{'}$ & 7, 1 & 3, 0\tabularnewline
\hline 
type: $\theta^{''}$ & 7, 0 & 3, 1\tabularnewline
\hline 
\end{tabular}

\noindent \medskip{}
\end{center}

Against any receiver strategy, the two types $\theta^{'}$ and $\theta^{''}$
get the same payoffs from $s^{''}$, but $\theta^{''}$ gets strictly
higher payoffs than $\theta^{'}$ from $s^{'}$. So, $\theta^{'}\succsim_{s^{''}}\theta^{''}$.

Consider now the Nash equilibrium in which the types pool on $s^{'}$,
i.e. $\pi_{1}^{*}(s^{'}|\theta^{'})=\pi_{1}^{*}(s^{'}|\theta^{''})=1,$
$\pi_{2}^{*}(a^{'}|s^{'})=1$, and $\pi_{2}^{*}(a^{''}|s^{''})=1$.
It passes the Intuitive Criterion since the off-path signal $s^{''}$
is not equilibrium dominated for either type. On the other hand, RCE
requires that every action played with positive probability in $\pi_{2}^{*}(\cdot|s^{''})$
best responds to some belief $p$ about sender's type satisfying $\frac{p(\theta^{''})}{p(\theta^{'})}\le\frac{\lambda(\theta^{''})}{\lambda(\theta^{'})}=\frac{1}{3}$.
But action $a^{''}$ does not best respond to any such belief, so
$\pi^{*}$ is not an RCE. \hfill{} $\blacklozenge$
\end{example}

\subsection{\label{subsec:Divine-Equilibrium}Divine Equilibrium}

Next, we compare divine equilibrium with RCE and uRCE. For a strategy
profile $\pi^{*}$, let 
\[
D(\theta,s;\pi^{*})\coloneqq\{\alpha\in\text{MBR}(s)\text{ s.t. }\mathbb{E}_{\pi^{*}}[u_{1}\mid\theta]<u_{1}(\theta,s,\alpha)\}
\]
be the subset of mixed best responses\footnote{To be precise, $\text{MBR}(p,s):=\underset{\alpha\in\Delta(A)}{\arg\max}(\mathbb{E}_{\theta\sim p}[u_{2}(\theta,s,\alpha)])$
and $\text{MBR}(s):=\cup_{p\in\Delta(\Theta)}\text{MBR}(p,s)$.} to $s$ that would make type $\theta$ strictly prefer deviating
from the strategy $\pi_{1}^{*}(\cdot\mid\theta)$. Similarly let
\[
D^{\circ}(\theta,s;\pi^{*})\coloneqq\{\alpha\in\text{MBR}(s)\text{ s.t. }\mathbb{E}_{\pi^{*}}[u_{1}\mid\theta]=u_{1}(\theta,s,\alpha)\}
\]
be the set of mixed best responses that would make $\theta$ indifferent
to deviating.

Intuitively, RCE and uRCE are ``close'' to divine equilibrium because
both the definition of $\succsim_{s}$ and the divine equilibrium
belief restriction involve a condition of the form ``any receiver
play that makes one type weakly prefer $s$ must make another type
strictly prefer $s$.'' Propositions \ref{prop:divine_RCE} and \ref{prop:NWBR}
make this relationship precise.
\begin{prop}
\label{prop:divine_RCE}
\begin{enumerate}
\item If $\pi^{*}$ is a Nash equilibrium where $s^{'}$ is off-path, and
$\theta^{'}\succsim_{s^{'}}\theta^{''},$ then $D(\theta^{''},s^{'};\pi^{*})\cup D^{\circ}(\theta^{''},s^{'};\pi^{*})\subseteq D(\theta^{'},s^{'};\pi^{*})$.
\item Every divine equilibrium is a RCE.
\end{enumerate}
\end{prop}
However, the converse is not true, as the following example illustrates.
\begin{example}
\label{exa:3_signals}Consider the following signaling game with two
types and three signals, with prior $\lambda(\theta_{1})=2/3$.

\noindent \begin{center}\medskip{}
\begin{tabular}{|c|c|c|}
\hline 
$s^{'}$ & $a^{'}$ & $a^{''}$\tabularnewline
\hline 
\hline 
$\theta^{'}$ & 0, 1 & -1, 0\tabularnewline
\hline 
$\theta^{''}$ & 0, 0 & -1, 1\tabularnewline
\hline 
\end{tabular}$\quad$ %
\begin{tabular}{|c|c|c|}
\hline 
$s^{''}$ & $a^{'}$ & $a^{''}$\tabularnewline
\hline 
\hline 
$\theta^{'}$ & 2, 1 & -1, 0\tabularnewline
\hline 
$\theta^{''}$ & 1, 0 & -1, 1\tabularnewline
\hline 
\end{tabular}$\quad$ %
\begin{tabular}{|c|c|c|}
\hline 
$s^{'''}$ & $a^{'}$ & $a^{''}$\tabularnewline
\hline 
\hline 
$\theta^{'}$ & 5, 0 & -3, 1\tabularnewline
\hline 
$\theta^{''}$ & 0, 1 & -2, 0\tabularnewline
\hline 
\end{tabular}\medskip{}
\end{center}

We check that the following is a pure-strategy RCE: $\pi_{1}(s^{'}|\theta^{'})=\pi_{1}(s^{'}|\theta^{''})=1,\pi_{2}(a^{'}|s^{'})=1,\pi_{2}(a^{''}|s^{''})=1,\pi_{2}(a^{''}|s^{'''})=1.$
Evidently $\pi$ is a Nash equilibrium and no type is equilibrium-dominated
at any off-path signal. We now check that we do not have $\theta{}^{'}\succsim_{s^{''}}\theta^{''}$
or $\theta^{''}\succsim_{s^{'''}}\theta^{'}$. Observe that against
the receiver strategy $\tilde{\pi}_{2}(a^{'}|s)=\frac{1}{2}$ for
every $s$, $s^{''}$ is strictly optimal for $\theta^{''}$ but $s^{'''}$
is strictly optimal for $\theta^{'}$, so $\theta^{'}\not\succsim_{s^{''}}\theta^{''}$.
And for the receiver strategy $\hat{\pi}_{2}(a^{'}|s)=1$ for every
$s$, $s^{'''}$ is strictly optimal for $\theta^{'}$ but $s^{''}$
is strictly optimal for $\theta^{''}$, so $\theta^{''}\not\succsim_{s^{'''}}\theta^{'}$.
This shows the strategy profile is an RCE.

However, $D(\theta^{''},s^{''};\pi)\cup D^{\circ}(\theta^{''},s^{''};\pi)$
is the set of distributions on $\{a^{'},a^{''}\}$ that put at least
weight 0.5 on $a^{'}$. Any such distribution is in $D(\theta^{'},s^{''};\pi)$.
So in every divine equilibrium, the receiver plays a best response
to a belief that puts weight no less than 2/3 on $\theta^{'}$ after
signal $s^{''}$, which can only be $a^{'}$.\emph{}\footnote{As noted by \citet{van1987stability}, it may seem more natural to
replace the set $\alpha\in\text{MBR}(m)$ in the definitions of $D$
and $D^{0}$ with the larger set $\alpha\in\text{co(BR}(s)),$ which
leads to the weaker equilibrium refinement that \citet*{sobel1990fixed}
call ``co-divinity''. This example also shows that RCE need not
be co-divine.} \hfill{} $\blacklozenge$
\end{example}
This example illustrates one difference between divine equilibrium
and RCE: Under divine equilibrium, the beliefs after signal $s^{''}$
only depend on the comparison between the payoffs to $s^{''}$ with
those of the equilibrium signal $s^{'}$, while the compatibility
criterion also considers the payoffs to a third signal $s^{'''}.$
In the learning model, this corresponds to the possibility that $\theta^{'}$
chooses to experiment with $s^{'''}$ at beliefs that induce $\theta^{''}$
to experiment with $s^{''}.$

RCE differs from divine equilibrium in another way, as divine equilibrium
involves an \emph{iterative} application of a belief restriction.
The next example illustrates this difference\footnote{We thank Joel Sobel for this example.}.
\begin{example}
\label{exa:vs_divine_iterated} There are three types, $\theta^{'},\theta^{''},\theta^{'''}$,
all equally likely. The signal space is $S=\{s^{'},s^{''}\},$ and
the set of receiver actions is $A=\{a^{1},a^{2},a^{3},a^{4}\}$. When
any sender type chooses the signal $s^{'},$ all parties get a payoff
of 0 regardless of the receiver's action. When the sender chooses
$s^{''}$, the payoffs are determined by the following matrix.

\noindent \begin{center}

\noindent %
\begin{tabular}{|c|c|c|c|c|}
\hline 
$s^{''}$ & $a^{1}$ & $a^{2}$ & $a^{3}$ & $a^{4}$\tabularnewline
\hline 
\hline 
$\theta^{'}$ & 1, 0.9 & -1, 0 & -2, 0 & -7, 0\tabularnewline
\hline 
$\theta^{''}$ & 5, 0 & 3, 1 & -1, 0 & -5, 0.8\tabularnewline
\hline 
$\theta^{'''}$ & -3, 0 & 5, 0 & 1, 1.7 & -3, 0.8\tabularnewline
\hline 
\end{tabular}

\noindent \end{center}

Consider the pure strategy profile $\pi_{1}^{*}(s^{'}|\theta)=1$
for all $\theta\in\Theta$ and $\pi_{2}^{*}(a^{4}|s)=1$ for all $s\in S$.
Since $\theta^{''}$ gains more from deviating to $s^{''}$ than $\theta^{'}$
does, applying the divine belief restriction for the off-path signal
$s^{''}$ eliminates the action $a^{1}$, since it is not a best response
to any belief $p\in\Delta(\Theta)$ with $p(\theta^{''})\ge p(\theta^{'})$.
But after action $a^{1}$ is deleted for the receiver after signal
$s^{''},$ type $\theta^{'''}$ now gains more from deviating to $s^{''}$
than $\theta^{''}$ does. So, applying the divine belief restriction
again eliminates actions $a^{2}$ and $a^{4}$, since it is not a
best response against any $p\in\Delta(\Theta)$ with $p(\theta^{'})=0$
(for now $s^{''}$ is equilibrium dominated for $\theta^{'})$ and
$p(\theta^{'''})\ge p(\theta^{''}).$ So $\pi^{*}$ is not a divine
equilibrium.

On the other hand, no type is equilibrium dominated at $s^{''}$ and
the only rational compatibility order is $\theta^{''}\succsim_{s^{''}}\theta^{'}$.
But $a^{4}$ is a best response against the belief $p(\theta^{'})=0,$
$p(\theta^{''})=0.6,$ $p(\theta^{'''})=0.4$, which belongs to the
set $\Delta(\Theta_{s^{''}})\bigcap P_{\theta^{''}\triangleright\theta^{'}}$.
So $\pi^{*}$ is an RCE. \hfill{} $\blacklozenge$
\end{example}
Finally, we show that every uRCE is path-equivalent to an equilibrium
that is not ruled out by the ``NWBR in signaling games'' test \citep{banks_equilibrium_1987,cho_signaling_1987},\footnote{This is closely related to, but not the same as, the NWBR property
of \citet{kohlberg_strategic_1986}.} which comes from iterative applications of the following pruning
procedure: after signal $s$ the receiver is required to put 0 probability
on those types $\theta$ such that
\[
D^{\circ}(\theta,s;\pi^{*})\subseteq\cup_{\theta^{'}\ne\theta}D(\theta^{'},s;\pi^{*}).
\]
If this would delete every type, then the procedure instead puts no
restriction on receiver's beliefs and no type is deleted.

By ``path-equivalent'' we mean that by modifying some of the receiver's
off-path responses, but without altering the sender's strategy or
the receiver's on-path responses, we can change the uRCE into another
uRCE that passes the NWBR test. Since every equilibrium passing the
NWBR test is universally divine\footnote{Universal divinity is defined as the iterative application of the
following procedure: after signal $s$ the receiver is required to
put 0 probability on those types $\theta$ such that
\[
D^{\circ}(\theta,s;\pi^{*})\cup D(\theta,s;\pi^{*})\subseteq\cup_{\theta^{'}\ne\theta}D(\theta^{'},s;\pi^{*}).
\]
} \citep{cho_signaling_1987}, this implies that every uRCE is path-equivalent
to a universally divine equilibrium.
\begin{prop}
\label{prop:NWBR}Every uRCE is path-equivalent to a uRCE that passes
the NWBR test.
\end{prop}
\begin{cor}
\label{cor:UGE_and_divine}Every uRCE is path-equivalent to a universally
divine equilibrium.
\end{cor}

\subsection{Summary}

To summarize this subsection, we note that for strategy profiles that
are on-path strict for the receiver, we have the following inclusion
relationships. The first inclusion should be understood as inclusion
up to path-equivalence. We use the symbol ``$\subsetneq$'' to mean
that the former solution set is always nested within the latter one
in every signaling game, and that there exist games where the nesting
relationship is strict. 
\[
\text{uRCE}\subsetneq\text{universally divine}\subsetneq\text{divine}\subsetneq\text{\text{RCE}}\subsetneq\text{Intuitive Criterion}\subsetneq\text{Nash}.
\]

In interpreting these inclusions, it is important to remember that
a universally divine equilibrium generically exists. In contrast,
we showed in Example \ref{exa:beer-quiche} that uRCE can fail to
exist for an open set of signaling games.\footnote{That is, for an open set of payoff vectors at the terminal nodes of
the game.}

\section{\label{sec:learning_model}Steady-State Learning in Signaling Games}

\subsection{Random Matching and Aggregate Play \label{subsec:learning11}}

We study the same discrete-time steady-state learning model as \citet{fudenberg_he_2017}
except for a different restriction on the players' prior beliefs over
other players' strategies.

There is a continuum of agents in the society, with a unit mass of
receivers and $\lambda(\theta)$ mass of type $\theta$ senders. Each
population is further stratified by age, with a fraction $(1-\gamma)\cdot\gamma^{t}$
of each population age $t$ for $t=0,1,2,...$ At the end of each
period, each agent has probability $0\le\gamma<1$ of surviving into
the next period, increasing their age by 1. With complementary probability,
the agent dies. Each agent's survival is independent of calendar time
and independent of the survival of other agents. At the start of the
next period, $(1-\gamma)$ new receivers and $\lambda(\theta)(1-\gamma)$
new type $\theta$ senders are born into the society, thus preserving
population sizes and the age distribution.

Agents play the signaling game every period against a randomly matched
opponent. Each sender has probability $(1-\gamma)\gamma^{t}$ of matching
with a receiver of age $t$, while each receiver has probability $\lambda(\theta)(1-\gamma)\gamma^{t}$
of matching with a type $\theta$ sender of age $t.$

\subsection{Learning by Individual Agents with Payoff Knowledge\label{subsec:learning22}}

Each agent is born into a player role in the signaling game: either
a receiver or a type $\theta$ sender. Agents know their role, which
is fixed for life. The agents' payoff each period is determined by
the outcome of the signaling game they played, which consists of the
sender's type, the signal sent, and the action played in response.
The agents observe this outcome, but the sender does not observe how
her matched receiver would have played had she sent a different signal.

In addition to only surviving to the next period with probability
$0\le\gamma<1$, agents discount\footnote{We separately consider survival probability and patience so that we
may consider agents who are impatient relative to their expected lifespan.
Such agents experiment early in their life cycle, but spend most of
their life myopically best responding to their beliefs, which makes
our analysis more tractable.} future utility flows by $0\le\delta<1$ and seek to maximize expected
discounted utility. Letting $u_{t}$ represent the payoff $t$ periods
from today, each agent's objective function is $\mathbb{E}[\sum_{t=0}^{\infty}(\gamma\delta)^{t}\cdot u_{t}]$.
(Define $0^{0}:=1$, so that a myopic agent just maximizes current
period's expected payoff in every period.)

Agents believe they face a fixed but unknown distribution of opponents'
aggregate play, updating their beliefs at the end of every period
based on the outcome in their own game. Formally, each sender is born
with a prior density function over receivers' behavior strategies,
$g_{1}:\Pi_{2}\to\mathbb{R}_{+}$ . Similarly, each receiver is born
with a prior density over the senders' behavior strategies, $g_{2}:\Pi_{1}\to\mathbb{R}_{+}$.
We denote the marginal distribution of $g_{1}$ on signal $s$ as
$g_{1}^{(s)}:\Delta(A)\to\mathbb{R}_{+}$, so that $g_{1}^{(s)}(\pi_{2}(\cdot|s))$
is the density of the new senders' prior over how receivers respond
to signal $s$. Similarly, we denote the $\theta$ marginal of $g_{2}$
as $g_{2}^{(\theta)}:\Delta(S)\to\mathbb{R}_{+}$, so that $g_{2}^{(\theta)}(\pi_{1}(\cdot|\theta))$
is the new receivers' prior density over the signal choice of type
$\theta$. \renewcommand{\theenumi}{(\alph{enumi})}

We now state a regularity assumption on agents' priors that will be
maintained throughout.
\begin{defn}
\label{def:regular_prior} A prior $g=(g_{1},g_{2})$ is \textbf{regular}
if
\end{defn}
\begin{enumerate}
\item {[}\emph{independence}{]} $g_{1}(\pi_{2})=\underset{s\in S}{\prod}g_{1}^{(s)}(\pi_{2}(\cdot|s))$
and $g_{2}(\pi_{1})=\underset{\theta\in\Theta}{\prod}g_{2}^{(\theta)}(\pi_{1}(\cdot|\theta))$.
\item {[}\emph{payoff knowledge}{]} $g_{1}$ puts probability 1 on $\Pi_{2}^{\bullet}$
and $g_{2}$ puts probability 1 on $\Pi_{1}^{\bullet}$.
\item {[}$g_{1}$ \emph{non-doctrinaire}{]} $g_{1}$ is continuous and strictly
positive on the interior of $\Pi_{2}^{\bullet}.$
\item {[}$g_{2}$ \emph{nice}{]} For each type $\theta$$,$ there are positive
constants $\left(\alpha_{s}^{(\theta)}\right)_{s\in S}$ such that
\[
\pi_{1}(\cdot|\theta)\mapsto\frac{g_{2}^{(\theta)}(\pi_{1}(\cdot|\theta))}{\prod_{s\in S}\pi_{1}(s|\theta){}^{\alpha_{s}^{(\theta)}-1}}
\]
is uniformly continuous and bounded away from zero on the relative
interior of $\Pi_{\theta}^{\bullet}$, the set of rational behavior
strategies of type $\theta$.
\end{enumerate}
This assumption bears the same name as the regularity assumption in
\citet{fudenberg_he_2017}, and is identical except that agents now
know others' payoffs and others' rationality. In the learning model,
this payoff knowledge translates into a restriction on the supports
of the priors $g_{1},g_{2}$, reflecting a dogmatic belief that senders
will never play dominated signals and receivers will never play conditionally
dominated actions. (These beliefs are correct in the learning model.)

Even with payoff knowledge, the receiver's prior can assign positive
probability to ex-ante dominated sender strategies. For instance,
in the signaling game below,

\noindent \begin{center}

\includegraphics[scale=0.3]{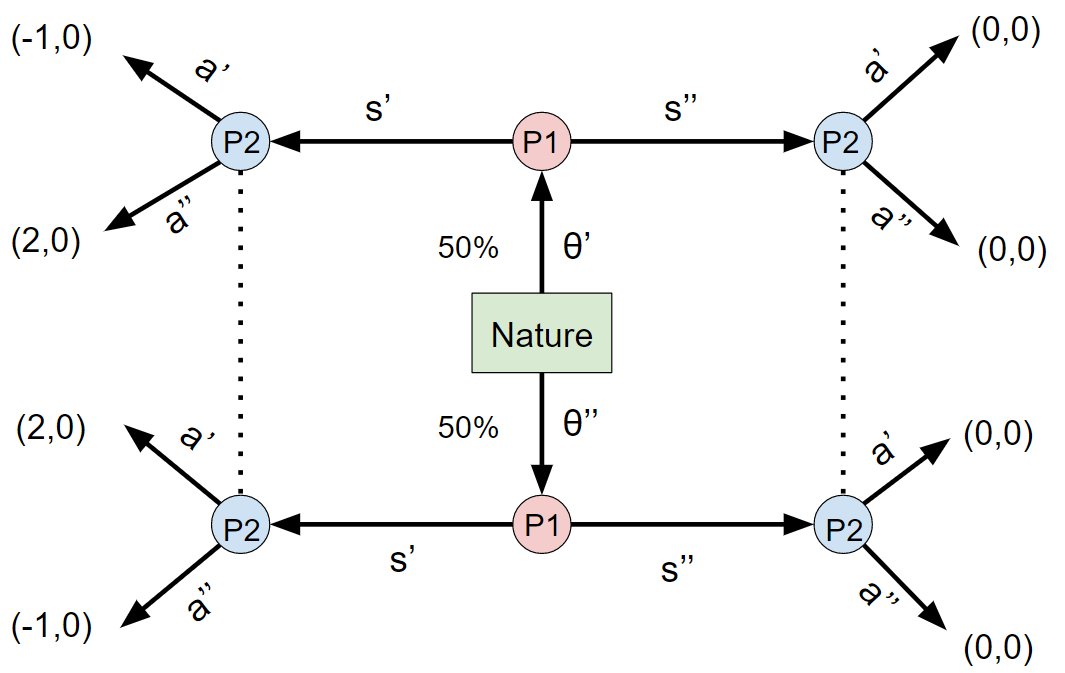}

\noindent \end{center} the sender strategy $\pi_{1}(s^{''}\mid\theta^{'})=\pi_{1}(s^{''}\mid\theta^{''})=1$
belongs to the set $\Pi_{1}^{\bullet}$, and so must belong to the
support of any regular receiver prior. But, even though $s^{''}\in S_{\theta^{'}}$
and $s^{''}\in S_{\theta^{''}}$, the receiver strategies to which
they respectively best respond form disjoint sets, and $\pi_{1}$
is ex-ante dominated because it is not a best response to any single
receiver strategy. It is nevertheless consistent for a receiver who
knows the sender's payoff as a function of their type to assign positive
density to $\pi_{1}$, because different types of agents can choose
best responses to different beliefs about receiver play.

\subsection{History and Aggregate Play \label{subsec:learning33}}

Let $Y_{\theta}[t]:=(\cup_{s\in S}(s\times A_{s}^{\text{BR}}))^{t}$
represent the set of possible histories for a type $\theta$ sender
with age $t$. Note that a valid history encodes the signal that $\theta$
sent each period and the (conditionally undominated) action that her
opponent played in response. Let $Y_{\theta}\coloneqq\bigcup_{t=0}^{\infty}Y_{\theta}[t]$
be the set of all histories for type $\theta$.

Similarly, write $Y_{2}[t]:=(\Theta\times S_{\theta})^{t}$ for the
set of possible histories for a receiver with age $t$. Each period,
his history encodes the type of the matched sender and the (undominated)
signal observed. The union $Y_{2}\coloneqq\bigcup_{t=0}^{\infty}Y_{2}[t]$
then stands for the set of all receiver histories.

The agents' dynamic optimization problems discussed in Subsection
\ref{subsec:learning22} give rise to\emph{ optimal policies}\footnote{For notational simplicity, we suppress the dependence of these optimal
policies on the effective discount factor $\delta\gamma$ and on the
priors.} $\sigma_{\theta}:Y_{\theta}\to S_{\theta}$ and $\sigma_{2}:Y_{2}\to\times_{s}(A_{s}^{\text{BR}})$.
Here, $\sigma_{\theta}(y_{\theta})$ is the signal that a type $\theta$
sender with history $y_{\theta}$ would send the next time she plays
the signaling game. Analogously, $\sigma_{2}(y_{2})$ is the pure
extensive-form strategy that a receiver with history $y_{2}$ would
commit to next time he plays the game. In the learning model, each
agent solves a (single-agent) dynamic optimization problem, and chooses
a deterministic optimal policy.

A \emph{state }$\psi$ of the learning model is a demographic description
of how many agents have each possible history. It can be viewed as
a distribution 
\[
\psi\in\left(\times_{\theta\in\Theta}\Delta(Y_{\theta})\right)\times\Delta(Y_{2}),
\]
and its components are denoted by $\psi_{\theta}\in\Delta(Y_{\theta})$
and $\psi_{2}\in\Delta(Y_{2})$.

Since each state $\psi$ is a distribution over histories and optimal
policies map histories to play, $\psi$ induces a distribution over
play (i.e., a rational behavior strategy) in the signaling game $\sigma(\psi)\in\Pi^{\bullet}$,
given by

\[
\sigma_{\theta}(\psi_{\theta})(s)\coloneqq\psi_{\theta}\left\{ y_{\theta}\in Y_{\theta}:\sigma_{\theta}(y_{\theta})=s\right\} 
\]
and

\[
\sigma_{2}(\psi_{2})(a\mid s)\coloneqq\psi_{2}\left\{ y_{2}\in Y_{2}:\sigma_{2}(y_{2})(s)=a\right\} .
\]

Here, $\sigma_{\theta}(\psi_{\theta})$ and $\sigma_{2}(\psi_{2})$
are the aggregate behaviors of the type $\theta$ and receiver populations
in state $\psi$, respectively. Note that the aggregate play of a
population can be stochastic even if the entire population uses the
same deterministic optimal policy, because different senders will
be matched with different receivers, and so different agents on the
same side will observe different histories and play differently.

Of particular interest are the \emph{steady states}, to be defined
more precisely in Section \ref{sec:Aggregate-Responses-and}. Loosely
speaking, a steady state induces a time-invariant distribution over
how the signaling game is played in the society.

\section{\label{sec:Aggregate-Responses-and}Aggregate Responses and Steady
State}

This section defines the notion of a steady state using the ``aggregate
responses'' of one population to the distribution of play in the
other. These responses are defined using the ``one-period forward''
maps that describe how the agents' policies induce a map from current
distributions over histories to what the distributions will be after
the agents are matched and play the game using the strategies their
policies prescribe.

\subsection{The Aggregate Sender Response \label{subsec:sender_APR}}

Fix the receivers' aggregate play at $\pi_{2}\in\Pi_{2}^{\bullet}$
and fix an optimal policy $\sigma_{\theta}$ for each type $\theta$.
The\emph{ one-period-forward map} \emph{for type $\theta$}, $f_{\theta}$,
describes the distribution over histories that will prevail next period
when the current distributions over histories in the type-$\theta$
population is $\psi_{\theta}$. The next definition specifies the
probability that $f_{\theta}[\psi_{\theta},\pi_{2}]$ assigns to the
history $(y_{\theta},(s,a))\in Y_{\theta}[t+1],$ that is to say a
one-period concatenation of $(s,a)$ onto the history $y_{\theta}\in Y_{\theta}[t]$.
\begin{defn}
The \emph{one-period-forward map for type $\theta$}, $f_{\theta}:\Delta(Y_{\theta})\times\Pi_{2}^{\bullet}\to\Delta(Y_{\theta})$
is 
\[
f_{\theta}[\psi_{\theta},\pi_{2}](y_{\theta},(s,a)):=\psi_{\theta}(y_{\theta})\cdot\gamma\cdot\boldsymbol{1}\{\sigma_{\theta}(y_{\theta})=s\}\cdot\pi_{2}(a\mid s)
\]
and $f_{\theta}(\varnothing):=1-\gamma$.
\end{defn}
To interpret, of the $\psi_{\theta}(y_{\theta})$ fraction of the
type-$\theta$ population with history $y_{\theta},$ a $\gamma$
fraction survives into the next period. The survivors all choose $\sigma_{\theta}(y_{\theta})$
next period, which is met with response $a$ with probability $\pi_{2}(a\mid\sigma_{\theta}(y_{\theta}))$.

Write $f_{\theta}^{T}$ for the $T$-fold application of $f_{\theta}$
on $\Delta(Y_{\theta}),$ holding fixed some $\pi_{2}$. It is easy
to show that $\lim_{T\to\infty}f_{\theta}^{T}(\psi_{\theta},\pi_{2})$
exists and is independent of the initial $\psi_{\theta}$. (This is
because for any two states $\psi_{\theta},\psi_{\theta}^{'}$, the
two distributions over histories $f_{\theta}^{T}(\psi_{\theta},\pi_{2})$
and $f_{\theta}^{T}(\psi_{\theta}^{'},\pi_{2})$ agree on all $Y_{\theta}[t]$
for $t<T$. As $T$ grows large, the two resulting distributions must
converge to each other since the fraction of very old agents with
very long histories is rare.) Denote this limit as $\tilde{\psi}_{\theta}^{\pi_{2}}.$
It is the distribution over type-$\theta$ history induced by the
receivers' aggregate play $\pi_{2}$.
\begin{defn}
The\emph{ aggregate sender response} $\mathscr{R}_{1}:\Pi_{2}^{\bullet}\to\Pi_{1}^{\bullet}$
is defined by 
\[
\mathscr{R}_{1}[\pi_{2}](s\mid\theta):=\tilde{\psi}_{\theta}^{\pi_{2}}(y_{\theta}:\sigma_{\theta}(y_{\theta})=s)
\]
\end{defn}
That is, $\mathscr{R}_{1}[\pi_{2}](\cdot\mid\theta)$ describes the
asymptotic aggregate play of the type-$\theta$ population when the
the aggregate play of the receiver population is fixed at $\pi_{2}$
each period. Note that $\mathscr{R}_{1}$ maps into $\Pi_{1}^{\bullet}$
because no type ever wants to send a dominated signal, even as an
experiment, regardless of their beliefs about the receiver's response.

Technically, $\mathscr{R}_{1}$ depends on $g_{1},\delta,$ and $\gamma$,
just like $\sigma_{\theta}$ does. When relevant, we will make these
dependencies clear by adding the appropriate parameters as superscripts
to $\mathscr{R}_{1}$, but we will mostly suppress them to lighten
notation.

\subsection{The Aggregate Receiver Response}

We now turn to the receivers, who have a passive learning problem.
They always observe the sender's type and signal at the end of each
period, so their optimal policy $\sigma_{2}$ myopically best responds
to the posterior belief at every history $y_{2}$.
\begin{defn}
The \emph{one-period-forward map for the receivers} $f_{2}:\Delta(Y_{2})\times\Pi_{1}^{\bullet}\to\Delta(Y_{2})$
is 
\[
f_{2}[\psi_{2},\pi_{1}](y_{2},(\theta,s)):=\psi_{2}(y_{2})\cdot\gamma\cdot\lambda(\theta)\cdot\pi_{1}(s|\theta)
\]
and $f_{2}(\varnothing):=1-\gamma$.
\end{defn}
As with the one-period-forward maps $f_{\theta}$ for senders, $f_{2}[\psi_{2},\pi_{1}]$
describes the distribution over receiver histories next period starting
with a society where the distribution is $\psi_{2}$ and the sender
population's aggregate play is $\pi_{1}.$ We write $\tilde{\psi}_{2}^{\pi_{1}}:=\lim_{T\to\infty}f_{2}^{T}(\psi_{2},\pi_{1})$
for the long-run distribution over $Y_{2}$ induced by fixing sender
population's play at $\pi_{1}$. (This limit is again independent
of the initial state $\psi_{2}.$)
\begin{defn}
The\emph{ aggregate receiver response} $\mathscr{R}_{2}:\Pi_{1}^{\bullet}\to\Pi_{2}^{\bullet}$
is 
\[
\mathscr{R}_{2}[\pi_{1}](a\mid s):=\tilde{\psi}_{2}^{\pi_{1}}(y_{2}:\sigma_{2}(y_{2})(s)=a)
\]
\end{defn}

\subsection{Steady States and Rational Patient Stability}

A \emph{steady-state strategy profile} is a pair of mutual aggregate
replies, so it is time-invariant under learning.\footnote{We focus on the steady states of the learning system, and do not study
convergence to steady states.}
\begin{defn}
$\pi^{*}$ is a \emph{steady-state strategy profile} if $\mathscr{R}_{1}^{g,\delta,\gamma}(\pi_{2}^{*})=\pi_{1}^{*}$
and $\mathscr{R}_{2}^{g,\delta,\gamma}(\pi_{1}^{*})=\pi_{2}^{*}$.
Denote the set of all such strategy profiles as $\Pi^{*}(g,\delta,\gamma)$.
\end{defn}
We now state two results about these steady states. We do not provide
a proof because they follow easily from analogous results in \citet{fudenberg_he_2017}.

First, steady-state profiles always exist.
\begin{prop}
\label{prop:existence}For any regular prior $g$ and any $0\le\delta,\gamma<1$,
$\Pi^{*}(g,\delta,\gamma)$ is non-empty and compact in the norm topology.
\end{prop}
The \emph{rationally patiently stable strategy profiles} correspond
to the set \emph{
\[
\lim_{\delta\to1}\lim_{\gamma\to1}\Pi^{*}(g,\delta,\gamma).
\]
}This order of limits was first introduced in \citet{fudenberg_steady_1993}.
It ensures agents spend most of their lifetime playing myopically
instead of experimenting, which is important for proving that rationally
patiently stable profiles are Nash equilibria.
\begin{defn}
For each $0\le\delta<1$, a strategy profile $\pi^{*}$ is \emph{$\delta$-stable
under $g$ }if there is a sequence $\gamma_{k}\to1$ and an associated
sequence of steady-state strategy profiles $\pi^{(k)}\in\Pi^{*}(g,\delta,\gamma_{k})$,
such that $\pi^{(k)}\to\pi^{*}$. Strategy profile $\pi^{*}$ is \emph{rationally
patiently stable under $g$ }if there is a sequence $\delta_{k}\to1$
and an associated sequence of strategy profiles $\pi^{(k)}$ where
each $\pi^{(k)}$ is $\delta_{k}$-stable under $g$ and $\pi^{(k)}\to\pi^{*}$.
Strategy profile $\pi^{*}$ is \emph{rationally patiently stable}
if it is rationally patiently stable under some regular prior $g$.
\end{defn}
Note that $\delta$-stable profiles always exist, since the space
of strategy profiles is compact so we can always extract a convergent
subsequence from a sequence of steady-state strategy profiles $\pi^{(k)}\in\Pi^{*}(g,\delta,\gamma_{k})$
with $\gamma_{k}\to1$. For the same reason, a rationally patiently
stable profile always exists.
\begin{prop}
\label{prop:nash} If strategy profile $\pi^{*}$ is rationally patiently
stable, then it is a Nash equilibrium.
\end{prop}
Note that Propositions \ref{prop:existence} and \ref{prop:nash}
apply even if all of the Nash equilibria of the game are in mixed
strategies; as noted above, the randomization here arises from the
random matching process.

\section{\label{sec:nec_and_suff}Rational Patient Stability, Payoff Knowledge,
and Equilibrium Refinements}

In this section, we relate the equilibrium refinements proposed in
Section \ref{sec:two_refinements} to the steady-state learning model.
We show that under certain strictness assumptions, RCE is necessary
for rational patient stability while uRCE is sufficient for rational
patient stability. We also discuss how payoff knowledge matters for
learning outcomes.

\subsection{RCE Is Necessary for Rational Patient Stability}

We show that any rationally patiently stable strategy profile satisfying
a strictness assumption must be an RCE. The key lemma is analogous
to Lemma 1 from \citet{fudenberg_he_2017}, so we will omit its proof.
\begin{lem}
\label{lem:sender_compatible} Suppose $\theta^{'}\succsim_{s}\theta^{''}$.
Then for any regular prior $g_{1}$, $0\le\delta,\gamma<1$, and any
$\pi_{2}\in\Pi_{2}^{\bullet}$, we have $\mathscr{R}_{1}[\pi_{2}](s\mid\theta^{'})\ge\mathscr{R}_{1}[\pi_{2}](s\mid\theta^{''})$.
\end{lem}
This result says over their lifetimes, the relative frequencies with
which different sender types experiment with signal $s$ respect the
rational compatible order $\succsim_{s}$. This follows from the fact
that sender types who are more compatible with a signal will play
it at least as often. The payoff knowledge embedded in $g_{1}$'s
support implies that senders never experiment in the hopes of seeing
a response which is highly profitable for the sender but dominated
for the receiver, such as the \textbf{X} action in Example \ref{exa:charity}
for $\theta_{\text{weak}}$. This extra assumption leads to a stronger
result than Lemma 1 from \citet{fudenberg_he_2017}, which is stated
in terms of the less-complete compatibility order.

For a fixed strategy profile $\pi$ and on-path signal $s^{*}$, let
$\mathbb{E}_{\theta|\pi_{1},s^{*}}[u_{2}(\theta,s^{*},a)]$ denote
the receiver's expected utility from responding to $s^{*}$ with $a$,
where the expectation over the sender's type $\theta$ is taken with
respect to the posterior type distribution after signal $s^{*}$ given
the sender's strategy $\pi_{1}(\cdot\mid\theta)$.
\begin{defn}
A Nash equilibrium $\pi^{*}$ is \emph{on-path strict for the receiver}
if for every on-path signal $s^{*},$ $\pi_{2}(a^{*}\mid s^{*})=1$
for some $a^{*}\in A$ and $\mathbb{E}_{\theta|\pi_{1},s^{*}}[u_{2}(\theta,s^{*},a^{*})]>\max_{a\ne a^{*}}\mathbb{E}_{\theta|\pi_{1},s^{*}}[u_{2}(\theta,s^{*},a)]$.
\end{defn}
We call this condition ``on-path'' strict for the receiver because
we do not make assumptions about the receiver's incentives after off-path
signals. For generic payoffs, all pure-strategy equilibria will be
on-path strict for the receiver.
\begin{thm}
\label{thm:necessary} Every strategy profile that is rationally patiently
stable and on-path strict for the receiver is an RCE.
\end{thm}
RCE rules out two kinds of receiver beliefs after signal $s$: those
that assign non-zero probability to equilibrium-dominated sender types,
and those that violate the rational compatibility order. The restriction
on equilibrium dominated types uses the assumption that the receiver
has a strict best response to each on-path signal to put a lower bound
on how slowly aggregate receiver play at on-path signals converges
to its limit.\footnote{If the receiver mixes after some equilibrium signal $s$ for type
$\theta$, then our techniques for showing that $\theta$ does not
experiment very much with equilibrium dominated signals do not go
through, but we do not have a counterexample.} The fact that the receiver beliefs respect the rational compatibility
order comes from Lemma \ref{lem:sender_compatible}, which uses our
assumptions about prior $g$ to derive restrictions on the aggregate
sender response $\mathscr{R}_{1}$, and show that these are reflected
in the aggregate receiver response. The proof of Theorem \ref{thm:necessary}
closely follows the the analogous proof in \citet{fudenberg_he_2017}
and is omitted.

\subsection{Quasi-Strict uRCE is Sufficient for Rational Patient Stability}

We now prove our main result: as a partial converse to Theorem \ref{thm:necessary},
we show that under additional strictness conditions, every uRCE is
path-equivalent to a rationally patiently stable strategy profile.
\begin{defn}
A \emph{quasi-strict uRCE} $\pi^{*}$ is a uRCE that is on-path strict
for the receiver, strict for the sender (that is, there exists an
equilibrium signal $s^{*}$ for each type $\theta$ with $u_{1}(\theta,s^{*},\pi_{2}^{*}(\cdot|s^{*}))>\max_{s\ne s^{*}}u_{1}(\theta,s,\pi_{2}^{*}(\cdot|s))$,
so every type strictly prefers its equilibrium signal to any other),
and satisfies $\mathbb{E}_{\pi^{*}}[u_{1}\mid\theta]>u_{1}(\theta,s^{'},a)$
for all $\theta,$ all off-path signals $s^{'}$ and all $a\in\text{BR}(\hat{P}(s^{'}),s^{'})$.
\end{defn}
The last condition in the definition of quasi-strictness requires
that every best response to $\hat{P}(s^{'})$ strictly deters every
type from deviating to $s^{'},$ whenever $s^{'}$ is off-path. Every
uRCE satisfies the weaker version of this condition where ``strictly
deters'' is replaced with ``weakly deters.''
\begin{thm}
\label{thm:sufficient} If $\pi^{*}$ is a quasi-strict uRCE, then
it is path-equivalent to a rationally patiently stable strategy profile.
\end{thm}
Theorem \ref{thm:sufficient} provides a constructive argument for
an equilibrium being rationally patiently stable in signaling games
with multiple RCE, such as Example \ref{exa:education}. It follows
from three lemmas on $\mathscr{R}_{1}$ and $\mathscr{R}_{2}$ that
are stated and proved in the rest of this subsection. Indeed, the
theorem remains valid in any modified learning model where $\mathscr{R}_{1}$
and $\mathscr{R}_{2}$ satisfy the conclusions of these lemmas.

Recall that Example \ref{exa:beer-quiche} showed a signaling game
with a unique RCE that is not a uRCE. As noted before, a rationally
patiently stable profile always exists. By Theorem \ref{thm:necessary},
the unique RCE of the game must be rationally patiently stable. This
shows uRCE is not a necessary condition for rational patient stability.

\subsubsection{$\mathscr{R}_{1}$ under a confident prior}

The first lemma shows that under a suitable prior, the aggregate sender
response of the dynamic learning model approximates the sender's static
best response function when applied to certain receiver strategies,
namely strategies that are ``close'' to one inducing a unique optimal
signal for each sender type. The precise meaning of ``close'' that
we use treats on- and off-path responses differently, so it requires
some auxiliary definitions.
\begin{defn}
Let $\pi^{*}$ be a strategy profile where every type plays a pure
strategy and the receiver plays a pure action after each on-path signal.
Say $\pi^{*}$ \emph{induces a unique optimal signal for each sender
type} if 
\[
\mathbb{E}_{\pi^{*}}[u_{1}\mid\theta]>\max_{s\ne\pi_{1}^{*}(\theta)}u_{1}(\theta,s,\pi_{2}^{*}(\cdot|s))
\]
for every type $\theta$.
\end{defn}
Starting with a strategy profile $\pi^{*}$ that induces a unique
optimal signal for each sender type, define for each off-path $s$
in $\pi^{*}$ the set of receiver actions $\tilde{A}(s):=\left\{ a:\mathbb{E}_{\pi^{*}}[u_{1}\mid\theta]>u_{1}(\theta,s,a)\text{\ \ensuremath{\forall\theta}}\right\} $
that strictly deter every type from deviation. Because $\pi_{2}^{*}$
induces a unique optimal signal, each $\tilde{A}(s)$ must contain
at least one element in the support of $\pi_{2}^{*}(\cdot|s)$, but
could also contain other actions. It is clear that if $\pi_{2}^{*}$
were modified off-path by changing each $\pi_{2}^{*}(\cdot|s)$ to
be an arbitrary mixture over $\tilde{A}(s),$ then the resulting strategy
profile would continue to induce (the same) unique optimal signal
for each sender type.

For $\pi^{*}$ that induces a unique optimal signal for each sender
type, write $B_{2}^{\text{on}}(\pi^{*},\epsilon)$ for the elements
of $\Pi_{2}^{\bullet}$ no more than $\epsilon$ away from $\pi_{2}^{*}$
at the on-path signals in $\pi_{1}^{*}$, that is 
\[
B_{2}^{\text{on}}(\pi^{*},\epsilon):=\left\{ \pi_{2}\in\Pi_{2}^{\bullet}:|\pi_{2}(a|s)-\pi_{2}^{*}(a|s)|\le\epsilon,\forall a,\text{ on-path }s\text{ in }\pi^{*}\right\} .
\]
Similarly, define $B_{2}^{\text{off}}(\pi^{*},\epsilon)$ as the elements
of $\Pi_{2}^{\bullet}$ putting no more than $\epsilon$ probability
on actions outside of $\tilde{A}(s)$ after each off-path $s$, where
$\tilde{A}(s)$ is the set of actions that would deter every type
from deviating to $s$, as above.
\[
B_{2}^{\text{off}}(\pi^{*},\epsilon):=\left\{ \pi_{2}\in\Pi_{2}^{\bullet}:\pi_{2}(\tilde{A}(s)|s)\ge1-\epsilon,\forall\text{ off-path }s\text{ in }\pi^{*}\right\} .
\]

\begin{lem}
\label{lem:sender1} Suppose $\pi^{*}$ induces a unique optimal signal
for each sender type. Then there exists a regular prior $g_{1}$,
some $0<\epsilon_{\text{off}}<1$, and a function $\gamma(\delta,\epsilon)$
valued in $(0,1)$, such that for every $0<\delta<1$, $0<\epsilon<\epsilon_{\text{off}}$,
and $\gamma(\delta,\epsilon)<\gamma<1$, if $\pi_{2}\in B_{2}^{\text{on}}(\pi^{*},\epsilon)\cap B_{2}^{\text{off}}(\pi^{*},\epsilon_{\text{off}})$,
then $|\mathscr{R}_{1}^{g_{1},\delta,\gamma}[\pi_{2}](s|\theta)-\pi_{1}^{*}(s|\theta)|<\epsilon$
for every $\theta$ and $s$.
\end{lem}
Note that the same $\epsilon$ appears in the hypothesis $\pi_{2}\in B_{2}^{\text{on}}(\pi^{*},\epsilon)$
as in the conclusion. That is, the aggregate sender response gets
closer to $\pi_{1}^{*}$ as receivers' play gets closer to $\pi_{2}^{*}$.

The idea is to specify a sender prior $g_{1}$ that is highly confident
and correct about the receiver's response to on-path signals, and
is also confident that the receiver responds to each off-path signal
$s$ with actions in $\tilde{A}(s)$. Take a signal $s^{'}$ other
than the one that $\theta$ sends in $\pi_{1}^{*}$. If $\theta$
has not experimented much with $s^{'}$, then her belief is close
to the prior and she thinks deviation does not pay. If $\theta$ has
experimented a lot with $s^{'}$, then by the law of large numbers
her belief is likely to be concentrated in $\tilde{A}(s^{'})$, so
again she thinks deviation does not pay. Since the option value for
experimentation eventually goes to 0, at most histories all sender
types are playing a myopic best response to their beliefs, meaning
they will not deviate from $\pi_{1}^{*}$. The intuition is similar
to that of Lemmas 6.1 and 6.4 from \citet{fudenberg_superstition_2006},
which says that we can construct a highly concentrated and correct
prior so that in the steady state, most agents have correct beliefs
about opponents' play both on and one step off the equilibrium path.

This lemma requires the assumption that $\pi^{*}$ is strict for the
sender. If $s^{*}$ were only weakly optimal for $\theta$ in $\pi^{*}$,
there could be receiver strategies arbitrarily close to $\pi_{2}^{*}$
that make some other signal $s^{'}\ne s^{*}$ strictly optimal for
$\theta$. In that case, we cannot rule out that a non-negligible
fraction of the $\theta$ population will rationally play $s^{'}$
forever when the receiver population plays close to $\pi_{2}^{*}$.

\subsubsection{$\mathscr{R}_{2}$ and learning rational compatibility}

Let $C$ be the set of sender strategies that respect the rational
compatibility order, that is 
\[
C:=\{\pi_{1}\in\Pi_{1}^{\bullet}:\pi_{1}(s|\theta)\ge\pi_{1}(s|\theta^{'})\ \text{whenever }\theta\succsim_{s}\theta^{'}\}.
\]

The next lemma shows that there is a prior for the receivers so that
when the aggregate sender play is any strategy in $C$, almost all
receivers end up with beliefs consistent with the rational compatibility
order. This lemma is the main technical contribution of the paper
and enables us to provide a sufficient condition for rational patient
stability when the \emph{relative} frequencies of off-path experiments
matter.
\begin{lem}
\label{lem:receiver1} For each $\epsilon>0$, there exists a regular
receiver prior $g_{2}$ and $0<\underline{\gamma}<1$ so that for
any $\underline{\gamma}<\gamma<1$, $0<\delta<1$, and $\pi_{1}\in C$,
\[
\mathscr{R}_{2}^{g_{2},\delta,\gamma}[\pi_{1}](\text{BR}(\hat{P}(s),s)\mid s)\ge1-\epsilon
\]
for each signal $s$.
\end{lem}
The key step in the proof is constructing a prior belief for the receivers
so that when the senders' aggregate play is sufficiently close to
the target equilibrium, the receiver beliefs respect the compatibility
order. This step was not necessary in \citet{fudenberg_superstition_2006},
which is the only other paper that has given a sufficient condition
for rational patient stability in a class of games.\footnote{Their result guarantees that the receivers' beliefs about the frequency
of type $\theta$ sending signal $s$ is within $\epsilon$ of the
truth. This is not sufficient for purposes, because when signal $s$
has probability 0 under a given sender strategy, perturbing the strategy
of every type by up to $\epsilon$ can generate arbitrary off-path
beliefs about the sender's type.}

To prove Lemma \ref{lem:receiver1}, we construct a Dirichlet prior
$g_{2}$ so that for any $s$ such that $\theta^{'}\succsim_{s}\theta^{''}$,
$g_{2}$ assigns much greater prior weight to $\theta^{'}$ playing
$s$ than to $\theta^{''}$ playing $s$..\footnote{The Dirichlet prior is the conjugate prior to multinomial data, and
corresponds to the updating used in fictitious play \citep{fudenberg1993learning}.
It is readily verified that if each of $g_{1}^{(\theta)}$ and $g_{2}^{(s)}$
is Dirichlet and independent of the other components, then $g$ is
regular. In the proof, we work with Dirichlet priors since they give
tractable closed-form expressions for the posterior mean belief of
the opponent's strategy after a given history.} In the absence of data, the receiver strongly believes that the senders
are using strategies $\pi_{1}$ such that $p(\theta^{''}|s)/p(\theta^{'}|s)\le\lambda(\theta^{''})/\lambda(\theta^{'})$.
This strong prior belief can only be overturned by a very large number
of observations to the contrary. But because $\pi_{1}\in C$ respects
the rational compatibility order, if the receiver has a very large
number of observations of senders choosing $s$, the law of large
numbers implies this large sample is unlikely to lead the receiver
to have a belief outside of $\hat{P}(s)$. So we can ensure that with
high probability sufficiently long-lived receivers play a best response
to $\hat{P}(s)$ after the off-path $s$.

Finally, we state a lemma that says for any Dirichlet receiver prior,
when lifetimes are long enough, the aggregate receiver response approximates
the receiver's best response function on-path when applied to a sender
strategy that provides strict incentives after every on-path signal.
Write $B_{1}^{\text{on}}(\pi^{*},\epsilon)$ for the elements of $\Pi_{1}^{\bullet}$
where each type $\theta$ plays $\epsilon$-close to $\pi_{1}^{*}(\cdot|\theta)$
, that is 
\[
B_{1}^{\text{on}}(\pi^{*},\epsilon):=\left\{ \pi_{1}\in\Pi_{1}^{\bullet}:|\pi_{1}(s|\theta)-\pi_{1}^{*}(s|\theta)|\le\epsilon,\forall\theta,s\right\} .
\]

\begin{lem}
\label{lem:receiver2} Fix a strategy profile $\pi^{*}$ where the
receiver has strict incentives after every on-path signal. For each
regular Dirichlet receiver prior $g_{2}$, there exists $\epsilon_{1}>0$
and a function $\gamma(\epsilon)$ valued in $(0,1)$, so that whenever
$\pi_{1}\in B_{1}^{\text{on}}(\pi^{*},\epsilon_{1})$, $0<\delta<1,$
and $\gamma(\epsilon)<\gamma<1$, we have $\mathscr{R}_{2}^{g_{2},\delta,\gamma}[\pi_{1}](a|s)-\pi_{2}^{*}(a|s)|<\epsilon$
for every on-path signal $s$ in $\pi^{*}$ and $a$.
\end{lem}
The intuition is that when the aggregate sender strategy is close
to $\pi_{1}^{*}$, the law of large numbers implies that after each
signal that $\pi_{1}^{*}$ gives positive probability, a receiver
with enough data is likely to have a belief close to the Bayesian
belief assigned by $\pi_{1}^{*}$. Coupled with the fact that $\pi_{1}^{*}$
is on-path strict for the receiver, this lets us conclude that long-lived
receivers play $\pi_{2}^{*}(\cdot|s)$ after every on-path $s$ with
high probability.

\section{\label{subsec:Payoff-Information-and-Steady-State}Payoff Information
and Steady-State Learning}

We revisit the examples from Section \ref{subsec:knowing_utilities_examples}
and discuss how prior beliefs reflecting knowledge or ignorance of
payoff information lead to different implications for learning.

\subsection{Example \ref{exa:charity}}

In Example \ref{exa:charity}, it follows from Lemma \ref{lem:sender_compatible}
that for any $0\le\delta,\gamma<1$, any receiver play $\pi_{2}\in\Pi_{2}^{\bullet}$,
and any regular prior $g$, we have $\mathscr{R}_{1}^{g_{1}}[\pi_{2}](\textbf{In}\mid\theta_{\text{strong}})\ge\mathscr{R}_{1}^{g_{1}}[\pi_{2}](\textbf{In}\mid\theta_{\text{weak}})$.
In the absence of payoff information, we show that there exists a
full-support prior $g_{1}$ so that, fixing $\pi_{2}$ to always play
\textbf{Down}, we get $\mathscr{R}_{1}^{g_{1}}[\pi_{2}](\textbf{In}\mid\theta_{\text{strong}})\le\mathscr{R}_{1}^{g_{1}}[\pi_{2}](\textbf{In}\mid\theta_{\text{weak}})$
for any $0\le\delta,\gamma<1$, with strict inequality for an open
set of parameter values.

Underlying this is the fact that if the conditionally dominated response
\textbf{X} is removed from the game tree, then $\theta_{\text{strong}}$
will experiment more frequently with \textbf{In} than $\theta_{\text{weak}}$
does because $\theta_{\text{strong}}$ potentially has more to gain.
This story breaks down if senders do not know receivers' payoffs and
thus suspect that \textbf{X} might be used after \textbf{In}. We now
show that for some full-support prior beliefs, $\theta_{\text{weak}}$
experiments more with \textbf{In} than $\theta_{\text{strong}}$ does
under \emph{any} patience level when receivers always play \textbf{Down}.

Let $g_{1}^{(\boldsymbol{\text{In}})}$ be Dirichlet with weights
$(1,K,1)$ on (\textbf{Up}, \textbf{Down}, \textbf{X}) for arbitrary
$K\ge4$. After observing $k\ge0$ instances of receivers responding
to \textbf{In} with \textbf{Down}, a sender would have the posterior
Dirichlet$(1,K+k,1)$. The $\theta_{\text{weak}}$ type's Gittins
index for \textbf{In} would be unchanged if her payoffs to (\textbf{Up},
\textbf{Down}, \textbf{X}) were $(3,-1,1)$ instead of $(1,-1,3)$,
by symmetry of her beliefs about \textbf{Up} and \textbf{X}. This
observation shows her Gittins index for \textbf{In} is at least as
large as $\theta_{\text{strong}}$'s, whose payoffs to (\textbf{Up},
\textbf{Down}, \textbf{X}) are $(2,-1,1)$. So the strong type switches
away from \textbf{In} after fewer observations of \textbf{Down} than
the weak type does (this includes the case of ``switching away''
after 0 observations of \textbf{Down}, i.e. the strong type never
experimenting with \textbf{In}.) We have proven $\mathscr{R}_{1}^{g_{1}}[\pi_{2}](\boldsymbol{\text{In}}\mid\theta_{\text{strong}})\le\mathscr{R}_{1}^{g_{1}}[\pi_{2}](\boldsymbol{\text{In}}\mid\theta_{\text{weak}})$
for any $0\le\delta,\gamma<1$.

Signal \textbf{In} is myopically suboptimal for both types, and by
the previous argument, the minimum effective discount factor $\delta\gamma$
that would induce at least one period of experimentation with \textbf{In}
is strictly higher for the strong type than the weak type. This shows
for an open set of $\delta,\gamma$ parameters, $\mathscr{R}_{1}^{g_{1}}[\pi_{2}](\textbf{In}\mid\theta_{\text{strong}})=0$
but $\mathscr{R}_{1}^{g_{1}}[\pi_{2}](\textbf{In}\mid\theta_{\text{weak}})>0$.

\subsection{Example \ref{exa:payoff_info_expands}}

In Example \ref{exa:payoff_info_expands} we showed that there is
an RCE in which the receivers play $a_{3}$ after \textbf{R}. Because
RCE is not a sufficient condition for rational patient stability,
this leaves open the question of whether this strategy can arise in
our learning model. Here we verify that it can, and also show that
``$a_{3}$ after \textbf{R}'' cannot be part of a patiently stable
outcome in the absence of payoff information. This is because patient
but inexperienced $\theta_{1}$'s without payoff information find
it plausible that receivers choose $a_{1}$ after \textbf{R}, so they
will experiment much more frequently with the off-path signal \textbf{R}
than $\theta_{2}$'s, for whom every possible response to \textbf{R}
leads to worse payoffs than their equilibrium payoff of 2. As a result,
receivers learn that \textbf{R}-senders have type $\theta_{1}$ so
they respond with $a_{2}$. On the other hand, when senders know ex-ante
that receivers will never choose $a_{1}$ after \textbf{R}, for some
priors there are steady states where no one ever experiments with
\textbf{R}. When this happens, the receivers' belief about the likelihood
ratio of the types following the off-path \textbf{R} is governed by
their initial beliefs about the distribution of sender play at the
start of learning, which may be arbitrary and thus support a richer
class of equilibrium profiles.

Specifically, in Example \ref{exa:payoff_info_expands}, suppose $g_{1}^{(\text{{\bf L}})}$
is $\text{Dirichlet }(1,10,1)$ over all three responses to \textbf{L},
while $g_{1}^{(\text{{\bf R}})}$ is $\text{Dirichlet}(1,1)$ on $A_{\text{{\bf R}}}^{\text{BR}}=\{a_{2},a_{3}\}$,
which reflects the sender's knowledge that $a_{1}$ is a conditionally
dominated response to \textbf{R}$.$ And suppose that $g_{2}^{\theta_{1}}$
is the $\text{Dirichlet}(2,1)$ distribution on $\{\text{\textbf{L}, \textbf{R}}\}$
and $g_{2}^{\theta_{2}}$ is the $\text{Dirichlet}(2,x)$ distribution,
where $x>0$ is a free parameter. For any $0\le\delta,\gamma<1,$
there exists a steady state where senders always choose \textbf{L}
and receivers always respond to \textbf{L} with $a_{2}$. This is
because the Gittins index for \textbf{R} is no larger than $-3$ for
$\theta_{1}$ and no larger than $1$ for $\theta_{2}$ after any
history, while the myopic expected payoff of \textbf{L} already exceeds
these values in the first period. The expected payoff of \textbf{L}
only increases with additional observations of $a_{2}$ after \textbf{L}.
On the receiver side, every positive-probability history $y_{2}$
must involve the senders playing \textbf{L} every period. Following
such a history, the receiver believes $\theta_{1}$ plays \textbf{L}
with probability at least $\frac{2}{3}$, hence an \textbf{L}-sender
is the $\theta_{1}$ type with probability at least $\frac{2/3}{1+2/3}=\frac{2}{5}$.
We have $a_{2}\in\text{BR}(\{p\},\text{{\bf L}})$ whenever $p(\theta_{1})\ge\frac{2}{5}$,
so we have shown that in the steady state receivers always play $a_{2}$
after \textbf{L}.

In this steady state, signal \textbf{R} is never sent, so by choosing
different values of $x>0$, we can sustain either $a_{2}$ or $a_{3}$
after \textbf{R} as part of a rationally patiently stable profile.
To be more precise, let $n_{1}$ and $n_{2}$ count the number of
times the two types of senders appear in a positive-probability history
$y_{2}$. The receiver's posterior assigns the following likelihood
ratio to the type of an \textbf{R}-sender: 
\[
\frac{1}{3+n_{1}}/\frac{x}{2+x+n_{2}}=\frac{1}{x}\cdot\left(\frac{2+x+n_{2}}{3+n_{1}}\right).
\]
Since the two types are equally likely, the fraction of receivers
with histories $y_{2}$ so that $0.9\le\left(\frac{2+x+n_{2}}{3+n_{1}}\right)\le1.1$
approaches 1 as $\gamma\to1.$ Depending on whether $x=1/4$ or $x=4$,
these receivers will play $a_{2}$ or $a_{3}$ after \textbf{R}, so
$\pi_{2}(a_{2}\mid\text{{\bf R}})=1$ and $\pi_{2}(a_{3}\mid{\bf R})=1$
are both $\delta$-stable for any $\delta\ge0$, under two different
regular priors reflecting payoff knowledge.

By contrast, Theorem 3 of \citet{fudenberg_he_2017} implies that
if priors $g_{1},g_{2}$ have full support on $\Pi_{2}$ and $\Pi_{1}$
respectively, then we must have $a_{2}$ after \textbf{R} in every
patiently stable profile. The idea is that when senders are patient
and long-lived, new $\theta_{1}$ start off by trying \textbf{R} but
new $\theta_{2}$ start off by trying \textbf{L}. When receivers play
$a_{2}$ after \textbf{L} with high probability, it is very unlikely
that $\theta_{2}$ ever switches away from \textbf{L}, providing a
bound on their frequency of playing \textbf{R}. On the other hand,
as their effective discount factor increases, $\theta_{1}$ will spend
arbitrarily many periods of its early life playing \textbf{R} in hopes
of getting the best payoff of 5, lacking the payoff knowledge that
$a_{1}$ is conditionally dominated for the receivers after \textbf{R}.
Receivers therefore end up learning that \textbf{R}-senders have type
$\theta_{1}$.

\section{Conclusion}

This paper studies non-equilibrium learning about other players' strategies
in the setting of signaling games. When the agents' prior beliefs
about their opponents' play reflect prior knowledge of others' payoff
functions, the steady states of societies of Bayesian learners can
be bounded by two equilibrium refinements, RCE and uRCE, that is we
get 
\[
\text{uRCE}\subseteq\text{rationally patiently stable profiles}\subseteq\text{\text{RCE}}.
\]
Furthermore, every divine equilibrium is an RCE. This is not true
for all of the equilibria that satisfy the Intuitive Criterion, which
suggests that divine equilibrium does a better job of capturing the
implications of learning. That said, we do not know the exact relationship
between rational patient stability and divine equilibrium, and there
is scope for sharpening our conclusions.

Divine equilibrium and RCE are only defined for signaling games. In
general extensive-form games, agents may find it optimal to play strictly
dominated strategies as experiments to learn about the consequences
of their other strategies, so requiring prior beliefs to be supported
on opponents' undominated strategies can lead to situations where
agents observe play that they had assigned zero prior probability.
We leave the associated complications for future work.

\bibliographystyle{ecta}
\bibliography{Gittins_eqm}

\begin{thebibliography}{14}
\newcommand{\enquote}[1]{``#1''}
\expandafter\ifx\csname natexlab\endcsname\relax\def\natexlab#1{#1}\fi

\bibitem[\protect\citeauthoryear{Banks and Sobel}{Banks and
  Sobel}{1987}]{banks_equilibrium_1987}
\textsc{Banks, J.~S. and J.~Sobel} (1987): \enquote{Equilibrium {Selection} in
  {Signaling} {Games},} \emph{Econometrica}, 55, 647--661.

\bibitem[\protect\citeauthoryear{Cho and Kreps}{Cho and
  Kreps}{1987}]{cho_signaling_1987}
\textsc{Cho, I.-K. and D.~M. Kreps} (1987): \enquote{Signaling {Games} and
  {Stable} {Equilibria},} \emph{Quarterly Journal of Economics}, 102, 179--221.

\bibitem[\protect\citeauthoryear{Dekel, Fudenberg, and Levine}{Dekel
  et~al.}{1999}]{dekel1999payoff}
\textsc{Dekel, E., D.~Fudenberg, and D.~K. Levine} (1999): \enquote{Payoff
  {Information} and {Self-Confirming} {Equilibrium},} \emph{Journal of Economic
  Theory}, 89, 165--185.

\bibitem[\protect\citeauthoryear{Esponda and Pouzo}{Esponda and
  Pouzo}{2016}]{esponda_berknash_2016}
\textsc{Esponda, I. and D.~Pouzo} (2016): \enquote{Berk-{Nash} {Equilibrium}:
  {A} {Framework} for {Modeling} {Agents} {With} {Misspecified} {Models},}
  \emph{Econometrica}, 84, 1093--1130.

\bibitem[\protect\citeauthoryear{Fudenberg and He}{Fudenberg and
  He}{2018}]{fudenberg_he_2017}
\textsc{Fudenberg, D. and K.~He} (2018): \enquote{Learning and Type
  Compatibility in Signaling Games,} \emph{Econometrica}, 86, 1215--1255.

\bibitem[\protect\citeauthoryear{Fudenberg and Kamada}{Fudenberg and
  Kamada}{2015}]{fudenberg_kamada_2015_rationalizable}
\textsc{Fudenberg, D. and Y.~Kamada} (2015): \enquote{Rationalizable
  partition-confirmed equilibrium,} \emph{Theoretical Economics}, 10, 775--806.

\bibitem[\protect\citeauthoryear{Fudenberg and Kamada}{Fudenberg and
  Kamada}{2018}]{fudenberg2018rationalizable}
---\hspace{-.1pt}---\hspace{-.1pt}--- (2018): \enquote{Rationalizable
  partition-confirmed equilibrium with heterogeneous beliefs,} \emph{Games and
  Economic Behavior}, 109, 364--381.

\bibitem[\protect\citeauthoryear{Fudenberg and Kreps}{Fudenberg and
  Kreps}{1993}]{fudenberg1993learning}
\textsc{Fudenberg, D. and D.~M. Kreps} (1993): \enquote{Learning {Mixed}
  {Equilibria},} \emph{Games and Economic Behavior}, 5, 320--367.

\bibitem[\protect\citeauthoryear{Fudenberg and Levine}{Fudenberg and
  Levine}{1993}]{fudenberg_steady_1993}
\textsc{Fudenberg, D. and D.~K. Levine} (1993): \enquote{Steady {State}
  {Learning} and {Nash} {Equilibrium},} \emph{Econometrica}, 61, 547--573.

\bibitem[\protect\citeauthoryear{Fudenberg and Levine}{Fudenberg and
  Levine}{2006}]{fudenberg_superstition_2006}
---\hspace{-.1pt}---\hspace{-.1pt}--- (2006): \enquote{Superstition and
  {Rational} {Learning},} \emph{American Economic Review}, 96, 630--651.

\bibitem[\protect\citeauthoryear{Kalai and Lehrer}{Kalai and
  Lehrer}{1993}]{kalai_rational_1993}
\textsc{Kalai, E. and E.~Lehrer} (1993): \enquote{Rational {Learning} {Leads}
  to {Nash} {Equilibrium},} \emph{Econometrica}, 61, 1019--1045.

\bibitem[\protect\citeauthoryear{Kohlberg and Mertens}{Kohlberg and
  Mertens}{1986}]{kohlberg_strategic_1986}
\textsc{Kohlberg, E. and J.-F. Mertens} (1986): \enquote{On the {Strategic}
  {Stability} of {Equilibria},} \emph{Econometrica}, 54, 1003--1037.

\bibitem[\protect\citeauthoryear{Sobel, Stole, and Zapater}{Sobel
  et~al.}{1990}]{sobel1990fixed}
\textsc{Sobel, J., L.~Stole, and I.~Zapater} (1990):
  \enquote{{Fixed-Equilibrium} {Rationalizability} in {Signaling} {Games},}
  \emph{Journal of Economic Theory}, 52, 304--331.

\bibitem[\protect\citeauthoryear{Van~Damme}{Van~Damme}{1987}]{van1987stability}
\textsc{Van~Damme, E.} (1987): \emph{Stability and Perfection of Nash
  Equilibria}, Springer-Verlag.

\end{thebibliography}

\appendix

\section{Appendix}

\subsection{Proof of Proposition \ref{prop:transitive_asymm}}
\begin{proof}
To show (1), suppose $\theta^{'}\succsim_{s^{'}}\theta^{''}$ and
$\theta^{''}\succsim_{s^{'}}\theta^{'''}.$ For any $\pi_{2}\in\Pi_{2}^{\bullet}$
where $s^{'}$ is weakly optimal for $\theta^{'''},$ it must be strictly
optimal for $\theta^{''},$ hence also strictly optimal for $\theta^{'}$.
This shows $\theta^{'}\succsim_{s^{'}}\theta^{'''}$.

To establish (2), partition the set of rational receiver strategies
as $\Pi_{2}^{\bullet}=\Pi_{2}^{+}\cup\Pi_{2}^{0}\cup\Pi_{2}^{-},$
where the three subsets refer to receiver strategies that make $s^{'}$
strictly better, indifferent, or strictly worse than the best alternative
signal for $\theta^{''}$. If the set $\Pi_{2}^{0}$ is nonempty,
then $\theta^{'}\succsim_{s^{'}}\theta^{''}$ implies $\theta^{''}\not\succsim_{s^{'}}\theta^{'}$.
This is because against any $\pi_{2}\in\Pi_{2}^{0}$, signal $s^{'}$
is strictly optimal for $\theta^{'}$ but only weakly optimal for
$\theta^{''}$. At the same time, if both $\Pi_{2}^{+}$ and $\Pi_{2}^{-}$
are nonempty, then $\Pi_{2}^{0}$ is nonempty. This is because both
$\pi_{2}\mapsto u_{1}(\theta^{''},s^{'},\pi_{2}(\cdot|s^{'}))$ and
$\pi_{2}\mapsto\max_{s^{''}\ne s^{'}}u_{1}(\theta^{''},s^{''},\pi_{2}(\cdot|s^{''}))$
are continuous functions, so for any $\pi_{2}^{+}\in\Pi_{2}^{+}$
and $\pi_{2}^{-}\in\Pi_{2}^{-},$ there exists $\alpha\in(0,1)$ so
that $\alpha\pi_{2}^{+}+(1-\alpha)\pi_{2}^{-}\in\Pi_{2}^{0}$. (Note
that $\pi_{2}^{+}$ and $\pi_{2}^{-}$ must be supported on $A_{s}^{\text{BR}}$
after every signal $s$, so the same must hold for the mixture $\alpha\pi_{2}^{+}+(1-\alpha)\pi_{2}^{-}.$
Thus, this mixture also belongs to $\Pi_{2}^{\bullet}.$) If only
$\Pi_{2}^{+}$ is nonempty and $\theta^{'}\succsim_{s^{'}}\theta^{''}$,
then $s^{'}$ is rationally strictly dominant for both $\theta^{'}$
and $\theta^{''}$. If only $\Pi_{2}^{-}$ is nonempty, then we can
have $\theta^{''}\succsim_{s^{'}}\theta^{'}$ only when $s^{'}$ is
never a weak best response for $\theta^{'}$ against any $\pi_{2}\in\Pi_{2}^{\bullet}$.
\end{proof}

\subsection{Proof of Proposition \ref{prop:uRCE_RCE_path}}
\begin{proof}
Let $\pi^{*}$ be a uRCE. We construct a path-equivalent RCE, $\pi^{\circ}$
as follows. Set $\pi_{1}^{\circ}=\pi_{1}^{*}$ and set $\pi_{2}^{\circ}(\cdot\mid s)=\pi_{2}^{*}(\cdot\mid s)$
for every on-path signal $s$.At each off-path signal $s$ where $\tilde{J}(s,\pi^{*})\ne\varnothing$,
let $\pi_{2}^{\circ}(\cdot\mid s)$ prescribe some best response to
a belief in $\tilde{P}(s,\pi^{*})$.At each off-path signal $s$ where
$\tilde{J}(s,\pi^{*})=\varnothing$, let $\pi_{2}^{\circ}(\cdot\mid s)$
prescribe some best response to a belief in $\Delta(\Theta_{s})$.

In this strategy profile, the receiver's play is a best response to
rationality-compatible beliefs after every off-path $s$ by construction,
and because the sender's play is the same as before the receiver is
still playing best responses to on-path signals.

Because the on-path play of the receivers did not change, no sender
type wishes to deviate to any on-path signal. Now we check that no
sender type wishes to deviate to any off-path signal. Consider first
off-path $s$ where $\tilde{J}(s,\pi^{*})\ne\varnothing$. Here we
have $\tilde{J}(s,\pi^{*})\subseteq\Theta_{s}$, which implies that
$\tilde{P}(s,\pi^{*})\subseteq\hat{P}(s)$. By the definition of uRCE,
$\pi_{2}^{\circ}(\cdot\mid s)$ must deter every type from deviating
to such $s.$ Finally, no sender type wishes to deviate to any $s$
where $\tilde{J}(s,\pi^{*})=\varnothing$, by the definition of equilibrium
dominance.
\end{proof}

\subsection{Proof of Proposition \ref{prop:compatible_and_dominance}}
\begin{proof}
Fix a $\pi_{1}$ with $\pi_{1}(s|\theta^{''})>0$ but $\pi_{1}(s|\theta^{'})<1$.
Because the space of rational receiver strategies $\Pi_{2}^{\bullet}$
is convex, it suffices to show there is no receiver strategy $\pi_{2}\in\Pi_{2}^{\bullet}$
such that $\pi_{1}$ is a best response to $\pi_{2}$ in the ex-ante
strategic form. If $\pi_{1}$ is an ex-ante best response, then it
needs to be at least weakly optimal for type $\theta^{''}$ to play
$s$ against $\pi_{2}$. By $\theta^{'}\succsim_{s}\theta^{''}$,
this implies $s$ is strictly optimal for type $\theta^{'}$. This
shows $\pi_{1}$ is not a best response to $\pi_{2}$, as the sender
can increase her ex-ante expected payoffs by playing $s$ with probability
1 when her type is $\theta^{'}$.
\end{proof}

\subsection{Proof of Proposition \ref{prop:IC}}
\begin{proof}
Suppose $\pi^{*}$ does not pass the Intuitive Criterion. Then there
exists a type $\theta$ and a signal $s^{'}$ such that 
\[
u_{1}(\theta;\pi^{*})<\min_{a\in\text{BR}(\Delta(\widetilde{J}(s^{'},\pi^{*})),s)}u_{1}(\theta,s^{'},a).
\]
If $\pi^{*}$ were an RCE, then we would have $\pi_{2}^{*}(\cdot|s^{'})\in\Delta(\text{BR}(\tilde{P}(s,\pi^{*}),s))$.
Since $\tilde{P}(s,\pi^{*})\subseteq\Delta(\widetilde{J}(s^{'},\pi^{*})),$
we have
\[
u_{1}(\theta;\pi^{*})<u_{1}(\theta,s^{'},\pi_{2}^{*}(\cdot|s^{'})).
\]
This means $\pi^{*}$ is not a Nash equilibrium, contradiction.
\end{proof}

\subsection{Proof of Proposition \ref{prop:divine_RCE}}
\begin{proof}
To show (a), note first that if $D(\theta^{''},s^{'};\pi^{*})\cup D^{\circ}(\theta^{''},s^{'};\pi^{*})=\varnothing$
the conclusion holds vacuously. If $D(\theta^{''},s^{'};\pi^{*})\cup D^{\circ}(\theta^{''},s^{'};\pi^{*})$is
not empty, take any $\alpha^{'}\in$$D(\theta^{''},s^{'};\pi^{*})\cup D^{\circ}(\theta^{''},s^{'};\pi^{*})$
and define $\pi_{2}^{'}\in\Pi_{2}^{\bullet}$ by $\pi_{2}^{'}(\cdot|s^{'})=\alpha^{'}$,
$\ensuremath{\pi_{2}^{'}(\cdot|s)=\pi_{2}^{*}(\cdot|s)}$ for $s\neq s^{'}$.
Then 
\[
u_{1}(\theta^{''};\pi^{*})=\max_{s\ne s^{'}}u_{1}(\theta^{''},s,\pi_{2}^{'}(\cdot|s))\leq u_{1}(\theta^{''},s^{'},\pi_{2}^{'}(\cdot|s^{'}))=u_{1}(\theta^{''},s^{'},\alpha^{'}),
\]
 and when $\theta^{'}\succsim_{s^{'}}\theta^{''},$ this implies that
\[
u_{1}(\theta^{'};\pi^{*})=\max_{s\ne s^{'}}u_{1}(\theta^{'},s,\pi_{2}^{'}(\cdot|s))<u_{1}(\theta^{'},s^{'},\pi_{2}^{'}(\cdot|s^{'}))=u_{1}(\theta^{'},s,\alpha^{'}).
\]
Hence $\alpha^{'}\in D(\theta^{'},s^{'};\pi^{*}).$

To show (b)\textbf{, }suppose $\pi^{*}$ is a divine equilibrium.
Then it is a Nash equilibrium, and furthermore for any off-path signal
$s^{'}$ where $\theta^{'}\succsim_{s^{'}}\theta^{''},$ Proposition
\ref{prop:divine_RCE}(a) implies that 
\[
D(\theta^{''},s^{'};\pi^{*})\cup D^{\circ}(\theta^{''},s^{'};\pi^{*})\subseteq D(\theta^{'},s^{'};\pi^{*}).
\]
Since $\pi^{*}$ is a divine equilibrium, $\pi_{2}^{*}(\cdot|s^{'})$
must then best respond to some belief $p\in\Delta(\Theta)$ with $\dfrac{p(\theta^{''})}{p(\theta^{'})}\le\dfrac{\lambda(\theta^{''})}{\lambda(\theta^{'})}$.
Considering all $(\theta^{'},\theta^{''})$ pairs, we see that in
a divine equilibrium $\pi_{2}^{*}(\cdot|s^{'})$ best responds to
some belief in 
\[
\bigcap_{(\theta^{'},\theta^{''})\text{ s.t. }\theta^{'}\succsim_{s^{'}}\theta^{''}}P_{\theta^{'}\triangleright\theta^{''}}.
\]
 At the same time, in every divine equilibrium, belief after off-path
$s^{'}$ puts zero probability on equilibrium-dominated types, meaning
$\pi_{2}^{*}(\cdot\mid s^{'})$ best responds $\Delta(\widetilde{J}(s^{'},\pi^{*}))$.
This shows $\pi^{*}$ is an RCE.
\end{proof}

\subsection{Proof of Proposition \ref{prop:NWBR}}
\begin{proof}
Consider a uRCE $\pi^{*}$. For every off-path $s$, perform the following
modifications on $\pi_{2}^{*}(\cdot|s)$: if the first-round application
of the NWBR procedure would have deleted every type, then do not modify
$\pi_{2}^{*}(\cdot|s)$. Otherwise, find some $\theta_{s}$ not deleted
by the iterated NWBR procedure, then change $\pi_{2}^{*}(\cdot|s)$
to some action in $\text{BR}(\{\theta_{s}\},s)$, i.e. a best response
to the belief putting probability 1 on $\theta_{s}$.

This modified strategy profile passes the NWBR test. We now establish
that it remains a uRCE by checking that for those off-path $s$ where
$\pi_{2}^{*}(\cdot|s)$ was modified, the modified version is still
a best response to $\hat{P}(s)$. (By uniformity, this would ensure
that the modified receiver play continues to deter every type from
deviating to $s$.)

Type $\theta_{s}$ satisfies $\theta_{s}\in\Theta_{s}$. Otherwise,
$D^{\circ}(\theta_{s},s;\pi^{*})=\varnothing$ and $\theta_{s}$ would
have been deleted by NWBR in the first round. Now it suffices to argue
there is no $\theta^{'}$ such that $\theta^{'}\succsim_{s}\theta_{s}$,
which implies the belief putting probability 1 on $\theta_{s}$ is
in $\hat{P}(s)$. If there were such $\theta^{'},$ by Proposition
\ref{prop:divine_RCE}(a) we would have $D^{\circ}(\theta_{s},s;\pi^{*})\subseteq D(\theta^{'},s;\pi^{*})$,
so $\theta_{s}$ should have been deleted by NWBR in the first round,
contradicting the fact that $\theta_{s}$ survives all iterations
of the NWBR procedure.
\end{proof}

\subsection{Proof of Corollary \ref{cor:UGE_and_divine}}
\begin{proof}
This is follows from Proposition \ref{prop:NWBR} because every NWBR
equilibrium is a universally divine equilibrium.
\end{proof}

\subsection{Proof of Lemma \ref{lem:sender1}}
\begin{proof}
Here are three lemmas from \citet{fudenberg_superstition_2006}:

\textbf{FL06 Lemma A.1}: Suppose $\{X_{k}\}$ is a sequence of i.i.d.
Bernoulli random variables with $\mathbb{E}[X_{k}]=\mu$, and define
for each $n$ the random variable

\[
S_{n}\coloneqq\frac{|\sum_{k=1}^{n}(X_{k}-\mu)|}{n}.
\]
Then for any $\underline{n},\bar{n}\in\mathbb{N}$, 
\[
\mathbb{P}\left[\max_{\underline{n}\le n\le\bar{n}}S_{n}>\epsilon\right]\le\frac{2^{7}}{3}\cdot\frac{1}{\underline{n}}\cdot\frac{\mu}{\epsilon^{4}}.
\]

\textbf{FL06 Lemma A.2}: For all $\epsilon,\epsilon'>0$, there is
an $N>0$ so that for all $\delta,\gamma,g,\pi$, signal $s$ and
action $a\in A$, 
\[
\psi_{\theta}^{\pi_{2};(g,\delta,\gamma)}\left\{ y_{\theta}:|\hat{\pi}_{2}(a|s;y_{\theta})-\pi_{2}(a|s)|>\epsilon,\#(s|y_{\theta})>N\right\} <\epsilon'.
\]
(Here, $\hat{\pi}_{2}(a|s;y_{\theta})$ is the empirical frequency
of receiver playing $a$ after signal $m$ in history $y_{\theta},$
that is to say $\hat{\pi}_{2}(a|s;y_{\theta})=\#((a,s),y_{\theta})/\#(s,y_{\theta})$.)

\textbf{FL06 Lemma A.4}: For all $\epsilon,\epsilon'>0$ and $\delta<1$,
there exists $N$ such that for all $\pi$, $g,$ and $\gamma$, we
get
\[
\psi_{\theta}^{\pi_{2};(g,\delta,\gamma)}\left\{ y_{\theta}\notin Y_{\theta}(\epsilon),\#(\sigma_{\theta}(y_{\theta}),y_{\theta})>N\right\} \le\epsilon'
\]
where $Y_{\theta}(\epsilon)\subseteq Y_{\theta}$ are those histories
$y_{\theta}$ where 
\[
\max_{s\in S}u_{1}(\theta,s|y_{\theta})\le u_{1}(\sigma_{\theta}(y_{\theta})|y_{\theta})+\epsilon,
\]
that is, type $\theta$ is playing a myopic $\epsilon$ best response
according to posterior belief after history $y_{\theta}$.

Now we proceed with our argument.

Since $\pi^{*}$ is strict on-path , there exist $\xi_{1},\xi_{2}>0$
such that whenever $\pi_{2}$ satisfies $|\pi_{2}(a|s)-\pi_{2}^{*}(a|s)|\le\xi_{1}$
for every on-path $s$ and action $a$, while for every off-path $s$
we have $\pi_{2}(\tilde{A}(s)|s)\ge1-\xi_{1}$, then for each type
$\theta$ we get 
\[
u_{1}(\theta,\pi_{1}^{*}(\theta),\pi_{2})>\xi_{2}+\max_{s\ne\pi_{1}^{*}(\theta)}u_{1}(\theta,s,\pi_{R}).
\]
That is, if receiver plays $\xi_{1}$-close to $\pi^{*}$ on-path
and $\xi_{1}$-close to $\tilde{A}(s)$ off-path, then for every type
of sender, playing the prescribed equilibrium signal is strictly better
than any other signal by at least $\xi_{2}>0$.

Following \citet{fudenberg_superstition_2006}, consider a prior $g_{1}$
such that whenever sender has fewer than $\underline{n}\coloneqq2^{11}/\xi_{1}^{4}$
observations of playing signal $s$, her belief as to receiver's probability
of taking action $a$ after signal $s$ differs from $\pi_{1}^{*}(a|s)$
by no more than $\xi_{1}$ if $s$ is on-path, while her belief as
to the probability that receiver strategy assigns to $\tilde{A}(s)$
is at least $1-\xi$ if $s$ is off-path. Also, let $\epsilon_{\text{off}}\coloneqq\xi_{1}/2$.

Now let $\delta\in(0,1)$ and $0<\epsilon<\epsilon_{\text{off}}$
be given. We construct $\gamma(\delta,\epsilon)$ satisfying the conclusion
of the lemma.

To do this, in FL06 Lemma A.4 put $\epsilon=\xi_{2}$ and $\epsilon^{'}=\epsilon/6$,
to obtain a $N_{1}(\epsilon)$. Next, in FL06 Lemma A.2 put $\epsilon=\xi_{1}/2$,
$\epsilon^{'}=\epsilon/6$, to obtain $N_{2}(\epsilon)$. Let $N(\epsilon)\coloneqq N_{1}(\epsilon)\vee N_{2}(\epsilon)$.
There are 5 classes of exceptional histories for type $\theta$ that
can lead to playing some signal $\hat{s}$ other than the one prescribed
by the equilibrium strategy, $s^{*}\coloneqq\pi_{1}^{*}(\theta)$.

\textbf{Exception 1}: $\theta$ has played $\hat{s}$ fewer than $N(\epsilon)$
times before, that is $\sigma_{\theta}(y_{\theta})=\hat{s}$ but $\#(\hat{s},y_{\theta})<N(\epsilon)$.
Such histories can be made to have mass no larger than $\epsilon/6$
by taking $\gamma(\delta,\epsilon)$ large enough.

\textbf{Exception 2}: $y_{\theta}$ is in the exceptional set described
in FL06 Lemma A.4. But by choice of $N(\epsilon)\ge N_{1}(\epsilon)$,
we know that
\[
\psi_{\theta}^{\pi_{2};(g,\delta,\gamma)}\left\{ y_{\theta}\notin Y_{\theta}(\xi_{2}),\#(\sigma_{\theta}(y_{\theta}),y_{\theta})>N(\epsilon)\right\} \le\epsilon/6.
\]
\textbf{Exception 3}: $\theta$ has played $\hat{s}$ more than $N(\epsilon)$
times, but has a misleading sample. By FL93 Lemma A.2, 
\[
\psi_{\theta}^{\pi_{2};(g,\delta,\gamma)}\left\{ y_{\theta}:|\hat{\pi}_{2}(a|\hat{s};y_{\theta})-\pi_{2}(a|\hat{s})|>\xi_{1}/2,\#(\hat{s}|y_{\theta})>N(\epsilon)\right\} <\epsilon/6.
\]
Since we have chosen $\pi\in B_{2}^{\text{on}}(\pi^{*},\epsilon)\cap B_{2}^{\text{off}}(\pi^{*},\epsilon_{\text{off}})$,
we know $\pi_{2}$ differs from $\pi_{2}^{*}$ by no more than $\epsilon_{\text{off}}=\xi_{1}/2$
after every on-path signal, and puts no more weight than $\xi_{1}/2$
on actions not in $\tilde{A}(s)$ after off-path signal $s$. So in
particular, 
\[
\psi_{\theta}^{\pi_{2};(g,\delta,\gamma)}\left\{ y_{\theta}:\begin{array}{c}
|\hat{\pi}_{2}(a|\hat{s};y_{\theta})-\pi_{2}^{*}(a|\hat{s})|>\xi_{1}\text{ if }\hat{s}\text{ on-path}\text{, or}\\
\hat{\pi}_{2}(\tilde{A}(\hat{s})|\hat{s})<1-\xi_{1}\text{ if }\hat{s}\text{ off-path}\\
\#(\hat{s}|y_{\theta})>N(\epsilon)
\end{array}\right\} <\epsilon/6.
\]
\textbf{Exception 4}: $\theta$ has played the equilibrium signal
$s^{*}$ more than $N(\epsilon)$ times, but has a misleading sample.
As before, we get
\[
\psi_{\theta}^{\pi_{2};(g,\delta,\gamma)}\left\{ y_{\theta}:|\hat{\pi}_{2}(a|s^{*};y_{\theta})-\pi_{2}^{*}(a|s^{*})|>\xi_{1},\#(s^{*}|y_{\theta})>N(\epsilon)\right\} <\epsilon/6.
\]
\textbf{Exception 5}: $\theta$ has played the equilibrium signal
$s^{*}$ between $\underline{n}$ and $N(\epsilon)$ times, but has
a misleading sample. Let $X_{k}\in\{0,1\}$ denote whether $\theta$
sees the equilibrium response $\pi_{2}^{*}(s^{*})$ the $k$-th time
she plays $s^{*}$ ($X_{k}=0$) or whether she sees instead a different
response $(X_{k}=1)$. As in FL06 Lemma A.1, define 
\[
S_{n}\coloneqq\frac{|\sum_{k=1}^{n}(X_{k}-\mu)|}{n}
\]
where $\mu=1-\pi_{2}(\pi_{2}^{*}(s^{*})|s^{*})<\epsilon$ since $s^{*}$
is an on-path signal in $\pi^{*}$.

The probability that the fraction of responses other than $\pi_{1}^{*}(s^{*})$
exceeds $\xi_{1}$ between the$\underline{n}$-th time and $N(\epsilon)$-th
time that $\theta$ plays $s^{*}$ is bounded above by FL06 Lemma
A.1, 
\begin{eqnarray*}
\mathbb{P}\left[\max_{\underline{n}\le n\le N(\epsilon)}S_{n}>\xi_{1}/2\right] & \le & \frac{2^{7}}{3}\cdot\frac{1}{\underline{n}}\cdot\frac{\mu}{(\xi_{1}/2)^{4}}\\
 & \le & \frac{1}{3}\cdot\mu\text{ (by choice of \ensuremath{\underline{n}})}\\
 & \le & \epsilon_{1}/3.
\end{eqnarray*}
Finally, at a history $y_{\theta}$ that does not belong to those
exceptions, we must have $\sigma_{\theta}(y_{\theta})=m^{*}$. This
is because $y_{\theta}$ is not in exception 1, so $\theta$ has played
$\sigma_{\theta}(y_{\theta})$ at least $N(\epsilon)$ times before,
and it is not in exception 2, so $\sigma_{\theta}(y_{\theta})$ is
a $\xi_{2}$ myopic best response to current beliefs. Yet the empirical
frequency for response after signal $\sigma_{\theta}(y_{\theta})$
is no more than $\xi_{1}$ away from $\pi_{2}^{*}(\sigma_{\theta}(y_{\theta}))$
as $y_{\theta}$ is not in exception 3 . Since the prior is Dirichlet
and also has this property, this means the current posterior belief
about response after signal $\sigma_{\theta}(y_{\theta})$ also has
this property. If $\#(s^{*},y_{\theta})>\underline{n}$, then $y_{\theta}$
not being in exceptions 4 or 5 implies belief as to response after
signal $s^{*}$ is also no more than $\xi_{1}$ away from $\pi_{2}^{*}(s^{*})$,
while if $\#(s^{*},y_{\theta})<\underline{n}$ then choice of prior
implies the same. In short, beliefs on both responses after $s^{*}$
and responses after $\sigma_{\theta}(y_{\theta})$ are no more than
$\xi_{1}$ away from their $\pi_{2}^{*}$ counterparts. But in that
case, no signal other than $s^{*}$ can be an $\xi_{2}$ best response.
\end{proof}

\subsection{Proof of Lemma \ref{lem:receiver1}\protect 
}\begin{proof}
For each $\xi>0$, consider the approximation to $P_{\theta^{'}\triangleright\theta^{''}}$,
\[
P_{\theta^{'}\triangleright\theta^{''}}^{\xi}:=\left\{ p\in\Delta(\Theta):\frac{p(\theta^{''})}{p(\theta^{'})}\le(1+\xi)\frac{\lambda(\theta^{''})}{\lambda(\theta^{'})}\right\} 
\]
and hence the approximation to $\hat{P}(s)$, 
\[
\hat{P}_{\xi}(s)\coloneqq\Delta(\Theta_{s^{'}})\bigcap\left\{ P_{\theta^{'}\triangleright\theta^{''}}^{\xi}:\theta^{'}\succsim_{s^{'}}\theta^{''}\right\} .
\]
Since the BR correspondence has a closed graph, there is an $\xi>0$
such that $\text{BR}(\hat{P}_{\xi}(s),s)=\text{BR}(\hat{P}(s),s)$.

Take some such $\xi.$ Next we will choose a series of constants.

\begin{itemize}
\item Pick $0<h<1$ such that $\frac{1-h}{1+h}>(1-\xi)^{1/3}.$
\item Pick $G>0$ such that for every $\theta\in\Theta$, $1/(h^{2}\cdot G\cdot(1-h)\cdot\lambda(\theta))<\epsilon/(4\cdot|S|\cdot|\Theta|^{2}).$
\item For each $\theta$, construct a Dirichlet prior on $S_{\theta}$ with
parameters $\alpha(\theta,s)\ge0$. Pick Dirichlet prior parameters
$\alpha(\theta,s)\ge0$ so that whenever $\theta\succsim_{s}\theta'$,
we have 
\begin{equation}
\alpha(\theta,s)-\alpha(\theta^{'},s)>(\sqrt{(4\cdot|S|\cdot|\Theta|^{2})/\epsilon}+1)\cdot G.\label{eq:prior_compatible}
\end{equation}
 In the event that $\theta\succsim_{s}\theta^{'}$ and $\theta^{'}\succsim_{s}\theta$,
put $\alpha(\theta,s)=\alpha(\theta^{'},s)$.
\item Pick $\underline{N}\in\mathbb{N}$ so that for any $N>\underline{N},$
$\theta,\theta^{'}\in\Theta$, we have 
\[
\mathbb{P}[(1-h)\cdot N\cdot\lambda(\theta)\le\text{Binom}(N,\lambda(\theta))\le(1+h)\cdot N\cdot\lambda(\theta)]>1-\frac{\epsilon}{4\cdot|\Theta|}
\]
and 
\[
\frac{(1-h)\cdot N\cdot\lambda(\theta^{'})}{(1+h)\cdot N\cdot\lambda(\theta)+\max_{\theta}\sum_{s\in S}\alpha(\theta,s)}>(1-\xi)^{1/3}\frac{\lambda(\theta^{'})}{\lambda(\theta)}.
\]
\item Pick $\underline{\gamma}\in(0,1)$ such that $1-(\underline{\gamma})^{\underline{N}+1}<\epsilon/4.$
\end{itemize}
Suppose the receiver's prior over the strategy of type$\theta$ is
Dirichlet with parameters $(\alpha(\theta,s))_{s\in S}$. We claim
that the conclusion of the lemma holds.

Fix some strategy $\pi_{1}\in C$. Write $\#(\theta|y_{2})$ for the
number of times the sender has been of $\theta$ type in history $y_{2}$,
while $\#(\theta,s|y_{2})$ counts the number of times type $\theta$
has sent signal $s$ in history $y_{2}$. Put $\psi_{2}=\psi_{2}^{\pi_{1};(g,\delta,\gamma)}$
and write $E\subseteq Y_{2}$ for those receiver histories with length
at least $\underline{N}$ satisfying 
\[
(1-h)\cdot N\cdot\lambda(\theta)\le\#(\theta|y_{2})\le(1+h)\cdot N\cdot\lambda(\theta)
\]
for every $\theta\in\Theta$. By the choice of $\underline{N}$ and
$\underline{\gamma}$, whenever $\gamma>\underline{\gamma}$ we have
$\psi(E)\ge1-\epsilon/2$. We now show that given $E$, the conditional
probability that the receiver's posterior belief after every off-equilibrium
signal $s$ lies in $\hat{P}_{\xi}(s)$ is at least $1-\epsilon/2$.
To do this, fix signal $s$ and two types with $\theta\succsim_{s}\theta^{'}$.

If $s$ is strictly dominated for both $\theta$ and $\theta^{'}$,
then according to the receivers' Dirichlet prior, $\theta$ and $\theta^{'}$
each sends $s$ with zero probability. Since $\pi\in\Pi_{1}^{\bullet},$
we have $\pi_{1}(s|\theta)=\pi_{1}(s|\theta^{'})=0$. So after every
positive-probability history, receiver's belief falls in $\hat{P}_{\xi}(s)$
as it puts zero probability on the $s$-sender being $\theta$ or
$\theta^{'}$. Henceforth we only consider the case where $s$ is
not strictly dominated for both.

After history $y_{2}$, the receiver's updated posterior likelihood
ratio for types $\theta$ and $\theta'$ upon seeing signal $s$ is
\begin{align*}
 & \frac{\lambda(\theta)}{\lambda(\theta')}\cdot\left(\frac{\alpha(\theta,s)+\#(\theta,s|y_{2})}{\#(\theta|y_{2})+\sum_{s\in S}\alpha(\theta,s)}/\frac{\alpha(\theta^{'},s)+\#(\theta^{'},s|y_{2})}{\#(\theta^{'}|y_{2})+\sum_{s\in S}\alpha(\theta^{'},s)}\right)\\
= & \frac{\lambda(\theta)}{\lambda(\theta')}\cdot\frac{\alpha(\theta,s)+\#(\theta,s|y_{2})}{\alpha(\theta^{'},s)+\#(\theta^{'},s|y_{2})}\cdot\frac{\#(\theta^{'}|y_{2})+\sum_{s\in S}\alpha(\theta^{'},s)}{\#(\theta|y_{2})+\sum_{s\in S}\alpha(\theta,s)}.
\end{align*}
Since we have $\#(\theta^{'}|y_{2})\ge(1-h)\cdot N\cdot\lambda(\theta^{'})$
while $\#(\theta|y_{2})\le(1+h)\cdot N\cdot\lambda(\theta)$, we get
\[
\frac{\#(\theta^{'}|y_{2})+\sum_{s\in S}\alpha(\theta^{'},s)}{\#(\theta|y_{2})+\sum_{s\in S}\alpha(\theta,s)}\ge\frac{(1-h)\cdot N\cdot\lambda(\theta^{'})}{(1+h)\cdot N\cdot\lambda(\theta)+\sum_{s\in S}\alpha(\theta,s)}>(1-\xi)^{1/3}\cdot\frac{\lambda(\theta')}{\lambda(\theta)}.
\]
If $s$ is strictly dominant for both $\theta$ and $\theta^{'}$,
then $\pi_{1}\in\Pi_{1}^{\bullet}$ means that $\pi_{1}(s|\theta)=\pi_{1}(s|\theta^{'})=1$.
In this case, $\#(\theta,s|y_{2})=\#(\theta|y_{2})$ and $\#(\theta^{'},s|y_{2})=\#(\theta^{'}|y_{2})$.
Since $\#(\theta|y_{2})\ge(1-h)\cdot N\cdot\lambda(\theta)$, $\#(\theta^{'}|y_{2})\le(1+h)\cdot N\cdot\lambda(\theta^{'})$,
we have:
\[
\frac{\alpha(\theta,s)+\#(\theta,s|y_{2})}{\alpha(\theta^{'},s)+\#(\theta^{'},s|y_{2})}\ge\frac{(1-h)\cdot N\cdot\lambda(\theta)}{\sum_{s\in S}\alpha(\theta^{'},s)+(1+h)\cdot N\cdot\lambda(\theta^{'})}\ge(1-\xi)^{1/3}\frac{\lambda(\theta)}{\lambda(\theta^{'})}.
\]
This shows the product is no smaller than $(1-\xi)^{2/3}\frac{\lambda(\theta)}{\lambda(\theta^{'})}$,
so receiver believes in $P_{\theta\triangleright\theta^{'}}^{\xi}$
after every history in $E$.

Now we analyze the term $\frac{\alpha(\theta,s)+\#(\theta,s|y_{2})}{\alpha(\theta^{'},s)+\#(\theta^{'},s|y_{2})}$
for the case where $s$ is not strictly dominant for both $\theta$
and $\theta^{'}$. We consider two cases, depending on whether $N$
is ``large enough'' so that the compatible type $\theta$ experiments
enough on average in a receiver history of length $N$ under sender
strategy $\pi_{1}$.

\textbf{Case A}: $\pi_{1}(s|\theta)\cdot N<G$. In this case, since
$\pi\in C$ and $\theta\succsim_{s}\theta'$, we must also have $\pi_{1}(s|\theta^{'})\cdot N<G$.
Then $\#(\theta',s|y_{2})$ is distributed as a binomial random variable
with mean smaller than $G$, hence standard deviation smaller than
$\sqrt{G}$. By Chebyshev's inequality, the probability that it exceeds
$(\sqrt{(4\cdot|S|\cdot|\Theta|^{2})/\epsilon}+1)\cdot G$ is no larger
than 
\[
\frac{1}{G\cdot(4\cdot|S|\cdot|\Theta|^{2})/\epsilon}<\frac{\epsilon}{4|S|\cdot|\Theta|^{2}}.
\]
But in any history $y_{2}$ where $\#(\theta',s|y_{R})$ does not
exceed this number, we would have 
\[
\alpha(\theta^{'},s)+\#(\theta^{'},s|y_{2})\le\alpha(\theta,s)\le\alpha(\theta,s)+\#(\theta,s|y_{2})
\]
by choice of the difference between prior parameters $\alpha(\theta',s)$
and $\alpha(\theta,s)$. Therefore $\frac{\alpha(\theta,s)+\#(\theta,s|y_{2})}{\alpha(\theta^{'},s)+\#(\theta^{'},s|y_{2})}\ge1$.
In summary, under Case A, there is probability no smaller than $1-\frac{\epsilon}{4|S|\cdot|\Theta|^{2}}$
that $\frac{\alpha(\theta,s)+\#(\theta,s|y_{2})}{\alpha(\theta^{'},s)+\#(\theta^{'},s|y_{2})}\ge1$.

\textbf{Case B}: $\pi_{1}(s|\theta)\cdot N\ge G$. In this case, we
can bound the probability that 
\[
\#(\theta,s|y_{2})/\#(\theta^{'},s|y_{2})\le\frac{\lambda(\theta)}{\lambda(\theta')}\cdot(\frac{1-h}{1+h})^{2}.
\]
Let $p\coloneqq\pi_{1}(s|\theta)$. Given that $\#(\theta|y_{2})\ge(1-h)\cdot N\cdot\lambda(\theta)$,
the distribution of $\#(\theta,s|y_{2})$ first order stochastically
dominates $\text{Binom}((1-h)\cdot N\cdot\lambda(\theta),p).$

On the other hand, given that $\#(\theta|y_{2})\le(1+h)\cdot N\cdot\lambda(\theta')$
and furthermore $\pi_{1}(s|\theta')\le\pi_{1}(s|\theta)=p$, the distribution
of $\#(\theta',s|y_{1})$ is first order stochastically dominated
by $\text{Binom}((1+h)\cdot N\cdot\lambda(\theta'),p).$

The first distribution has mean $(1-h)\cdot N\cdot\lambda(\theta)\cdot p$
with standard deviation no larger than $\sqrt{(1-h)\cdot N\cdot\lambda(\theta)\cdot p}$.
Thus 
\begin{align*}
 & \mathbb{P}\left[\text{Binom}((1-h)\cdot N\cdot\lambda(\theta),p)<(1-h)\cdot(1-h)\cdot N\cdot\lambda(\theta)\cdot p\right]\\
 & <1/(h\cdot\sqrt{p(1-h)N\lambda(\theta)})^{2}\le1/(h\cdot\sqrt{G\cdot(1-h)\cdot\lambda(\theta)})^{2}<\epsilon/(4\cdot|S|\cdot|\Theta|^{2})
\end{align*}
where we used the fact that $pN\ge G$ in the second-to-last inequality,
while the choice of $G$ ensured the final inequality.

At the same time, the second distribution has mean $(1+h)\cdot N\cdot\lambda(\theta')\cdot p$
with standard deviation no larger than $\sqrt{(1+h)\cdot N\cdot\lambda(\theta')\cdot p}$,
so 
\begin{align*}
 & \mathbb{P}\left[\text{Binom}((1+h)\cdot N\cdot\lambda(\theta'),p)>(1+h)\cdot(1+h)\cdot N\cdot\lambda(\theta')\cdot p\right]\\
 & <1/(h\cdot\sqrt{p(1+h)N\lambda(\theta')})^{2}\le1/(h\cdot\sqrt{G\cdot(1+h)\cdot\lambda(\theta')})^{2}<\epsilon/(4\cdot|S|\cdot|\Theta|^{2})
\end{align*}
by the same arguments. Combining the bounds on these two binomial
random variables, 
\[
\mathbb{P}\left[\frac{\text{Binom}((1-h)\cdot N\cdot\lambda(\theta),p)}{\text{Binom}((1+h)\cdot N\cdot\lambda(\theta'),p)}\le\frac{\lambda(\theta)}{\lambda(\theta')}\cdot(\frac{1-h}{1+h})^{2}\right]<\epsilon/(2\cdot|S|\cdot|\Theta|^{2}).
\]
Via stochastic dominance, this shows \emph{a fortiori}
\[
\mathbb{P}\left[\#(\theta,s|y_{2})/\#(\theta^{'},s|y_{2})\le\frac{\lambda(\theta)}{\lambda(\theta')}\cdot(\frac{1-h}{1+h})^{2}\right]<\epsilon/(2\cdot|S|\cdot|\Theta|^{2}).
\]
Therefore, for any $s,\theta,\theta'$ such that $\theta\succsim_{s}\theta'$,
\[
\psi\left(y_{2}:\frac{\alpha(\theta,s)+\#(\theta,s|y_{2})}{\alpha(\theta^{'},s)+\#(\theta^{'},s|y_{2})}\ge\frac{\lambda(\theta)}{\lambda(\theta')}\cdot(\frac{1-h}{1+h})^{2}\ |\ E\right)\ge1-\epsilon/(2\cdot|S|\cdot|\Theta|^{2}).
\]
This concludes case B.

In either case, at a history $y_{2}$ with $(1-h)\cdot N\cdot\lambda(\theta)\le\#(\theta|y_{2})\le(1+h)\cdot N\cdot\lambda(\theta)$
for every $\theta,$ for every pair $\theta,\theta'$ such that $\theta\succsim_{s}\theta'$,
we get $\frac{\alpha(\theta,s)+\#(\theta,s|y_{2})}{\alpha(\theta^{'},s)+\#(\theta^{'},s|y_{2})}\ge\frac{\lambda(\theta)}{\lambda(\theta')}\cdot(\frac{1-h}{1+h})^{2}$
with probability at least $1-\epsilon/(2\cdot|S|\cdot|\Theta|^{2})$.

But at any history $y_{2}$ where this happens, the receiver's posterior
likelihood ratio for types $\theta$ and $\theta'$ after signal $s$
satisfies
\begin{eqnarray*}
 &  & \frac{\lambda(\theta)}{\lambda(\theta')}\cdot\frac{\alpha(\theta,s)+\#(\theta,s|y_{2})}{\alpha(\theta^{'},s)+\#(\theta^{'},s|y_{2})}\cdot\frac{\#(\theta^{'}|y_{2})+\sum_{s\in S}\alpha(\theta^{'},s)}{\#(\theta|y_{2})+\sum_{s\in S}\alpha(\theta,s)}\\
 & \ge & \frac{\lambda(\theta)}{\lambda(\theta')}\cdot\frac{\lambda(\theta)}{\lambda(\theta')}\cdot\left(\frac{1-h}{1+h}\right){}^{2}\cdot(1-\xi)^{1/3}\cdot\frac{\lambda(\theta')}{\lambda(\theta)}\\
 & \ge & \frac{\lambda(\theta)}{\lambda(\theta')}\cdot(1-\xi)^{2/3}\cdot(1-\xi)^{1/3}\ge\frac{\lambda(\theta)}{\lambda(\theta')}\cdot(1-\xi).
\end{eqnarray*}
As there are at most $|\Theta|^{2}$ such pairs for each signal $s$
and $|S|$ total signals, 
\[
\psi\left(y_{2}:\begin{array}{c}
\frac{\lambda(\theta)}{\lambda(\theta')}\cdot\frac{\alpha(\theta,s)+\#(\theta,s|y_{2})}{\alpha(\theta^{'},s)+\#(\theta^{'},s|y_{2})}\cdot\frac{\#(\theta^{'}|y_{2})+\sum_{s\in S}\alpha(\theta^{'},s)}{\#(\theta|y_{2})+\sum_{s\in S}\alpha(\theta,s)}\\
\ge\frac{\lambda(\theta)}{\lambda(\theta')}\cdot(1-\xi)
\end{array}\ \forall s,\theta\succsim_{s}\theta'\ |E\right)\ge1-\epsilon/2
\]
as claimed. As the event $E$ has $\psi$-probability no smaller than
$1-\epsilon/2$, there is $\psi$ probability at least $1-\epsilon$
that receiver's posterior belief is in $\hat{P}_{\xi}(s)$ after every
off-path $s$.
\end{proof}

\subsection{Proof of Lemma \ref{lem:receiver2}}
\begin{proof}
Since $\pi^{*}$ is on-path strict for the receiver, there exists
some $\xi>0$ such that for every on-path signal $s$ and every belief
$p\in\Delta(\Theta)$ with
\begin{equation}
|p(\theta)-p(\theta;s,\pi^{*})|<\xi,\ \forall\theta\in\Theta\label{eq:approx_receiver_belief}
\end{equation}
(where $p(\cdot;s,\pi^{*})$ is the Bayesian belief after on-path
signal $s$ induced by the equilibrium $\pi^{*}$), we have $\text{BR}(p,s)=\{\pi_{2}^{*}(s)\}$.
For each $s,$ we show that there is a large enough $N(s,\epsilon)$
and small enough $\zeta(s)$ so that when receiver observes history
$y_{2}$ generated by any $\pi\in B_{\text{on}}(\pi^{*},\epsilon^{'})$
with $\epsilon^{'}<\zeta(s)/4$ and length at least $N(s,\epsilon)$,
there is probability at least $1-\frac{\epsilon}{2|S|}$ that receiver's
posterior belief satisfies (\ref{eq:approx_receiver_belief}). Hence,
conditional on having a history length of at least $N(s,\epsilon),$
there is $1-\frac{\epsilon}{2|S|}$ chance that receiver will play
as in $\pi_{2}^{*}$ after $s$. By taking the maximum $N^{*}(\epsilon)\coloneqq\max_{s}(N(s,\epsilon_{1}))$
and minimum $\epsilon_{1}\coloneqq\min_{s}\zeta(s)$, we see that
whenever history is length $N^{*}(\epsilon)$ or more, and $\pi\in B_{\text{on}}(\pi^{*},\epsilon^{'})$
with $\epsilon^{'}<\epsilon_{1}$, there is at least $1-\epsilon/2$
chance that the receiver's strategy matches $\pi_{2}^{*}$ after every
on-path signal . Since we can pick $\gamma(\epsilon)$ large enough
that $1-\epsilon/2$ measure of the receiver population is age $N^{*}(\epsilon)$
or older, we are done.

To construct $N(s,\epsilon)$ and $\zeta(s)$, let $\Lambda(s)\coloneqq\lambda\{\theta:\pi_{1}^{*}(s|\theta)=1\}$.
Find small enough $\zeta(s)\in(0,1)$ so that:
\begin{itemize}
\item $|\frac{\lambda(\theta)}{\Lambda(s)\cdot(1-\zeta(s))}-\frac{\lambda(\theta)}{\Lambda(s)}|<\xi$
\item $|\frac{\lambda(\theta)\cdot(1-\zeta(s))}{\Lambda(s)+(1-\Lambda(s))\cdot\zeta(s)}-\frac{\lambda(\theta)}{\Lambda(s)}|<\xi$
\item $\frac{\zeta(s)}{1-\zeta(s)}\cdot\frac{\lambda(\theta)}{\Lambda(s)}<\xi$
\end{itemize}
for every $\theta\in\Theta$. After a history $y_{2}$, the receiver's
posterior belief as to the type of sender who sends signal $s$ satisfies
\[
p(\theta|s;y_{2})\propto\lambda(\theta)\cdot\frac{\#(\theta,s|y_{2})+\alpha(\theta,s)}{\#(\theta|y_{2})+A(\theta)},
\]
where $\alpha(\theta,s)$ is the Dirichlet prior parameter on signal
$s$ for type $\theta$ and $A(\theta)\coloneqq\sum_{s\in S}\alpha(\theta,s)$.
By the law of large numbers, for long enough history length, we can
ensure that if $\pi_{1}(s|\theta)>1-\frac{\zeta(s)}{4},$ then 
\[
\frac{\#(\theta,s|y_{2})+\alpha(\theta,s)}{\#(\theta|y_{2})+A(\theta)}\ge1-\zeta(s)
\]
with probability at least $1-\frac{\epsilon}{2|S|^{2}}$, while if
$\pi_{1}(s|\theta)<\zeta(s)/4$, then 
\[
\frac{\#(\theta,s|y_{2})+\alpha(\theta,s)}{\#(\theta|y_{2})+A(\theta)}<\zeta(s)
\]
with probability at least $1-\frac{\epsilon}{2|S|^{2}}$. Moreover
there is some $N(s,\epsilon)$ so that there is probability at least
$1-\frac{\epsilon}{2|S|}$ that a history $y_{2}$ with length at
least $N(s,\epsilon)$ satisfies above for all $\theta$. But at such
a history, for any $\theta$ such that $\pi_{1}^{*}(s|\theta)=1$,
\[
p(\theta|s;y_{2})\ge\frac{\lambda(\theta)\cdot(1-\zeta(s))}{\Lambda(s)+(1-\Lambda(s))\cdot\zeta(s)}
\]
and
\[
p(\theta|s;y_{2})\le\frac{\lambda(\theta)}{\Lambda(s)\cdot(1-\zeta(s))},
\]
while for some $\theta$ such that $\pi_{1}^{*}(s|\theta)=0$, 
\[
p(\theta|s;y_{2})\le\frac{\zeta(s)}{1-\zeta(s)}\cdot\frac{\lambda(\theta)}{\Lambda(s)}.
\]

Therefore the belief $p(\cdot|s;y_{R})$ is no more than $\xi$ away
from $p(\theta;s,\pi^{*})$, as desired.
\end{proof}

\subsection{Proof of Theorem \ref{thm:sufficient}}
\begin{proof}
We will construct a regular prior $g$. We will then show that for
every $0<\delta<1$, there exists convex and compact sets of strategy
profiles $E_{j}\subseteq\Pi^{\bullet}$ with $E_{j}\downarrow E_{*}\subseteq B_{1}^{\text{on}}(\pi^{*},0)\cap B_{2}^{\text{on}}(\pi^{*},0)$
and a corresponding sequence of survival probabilities $\gamma_{j}\to1$
so that $(\mathscr{R}_{1}^{g,\delta,\gamma_{j}}[\pi_{2}],\mathscr{R}_{2}^{g,\delta,\gamma_{j}}[\pi_{1}])\in E_{j}$
whenever $\pi\in E_{j}$. We proved in \citet{fudenberg_he_2017}
that $\mathscr{R}_{1}$ and $\mathscr{R}_{2}$ are continuous maps,
so a fixed point theorem implies that for each $j$, some strategy
profile in $E_{j}$ is a steady state profile under parameters $(g,\delta,\gamma_{j})$.
Any convergent subsequence of these $j$-indexed steady state profiles
has a limit in $E_{*},$ so this limit agrees with $\pi^{*}$ on path.
This shows that for every $\delta$ there is a $\delta$-stable strategy
profile path-equivalent to $\pi^{*}$, so there is a rationally patiently
stable strategy profile with the same property.

\textbf{Step 1}: Constructing $g$ and some thresholds.

Since $\pi^{*}$ induces a unique optimal signal for each sender type,
by Lemma \ref{lem:sender1} find a regular sender prior $g_{1},$
$0<\epsilon_{\text{off}}<0$, and a function $\gamma_{\text{LM1}}(\delta,\epsilon)$.

In Lemma \ref{lem:receiver1}, substitute $\epsilon=\epsilon_{\text{off}}$
to find a regular receiver prior $g_{2}$ and $0<\underline{\gamma}_{\text{LM2}}<1$.

Finally, in Lemma \ref{lem:receiver2} let $g_{2}$ be as constructed
above to find $\epsilon_{\text{LM3}}>0$ and a function $\gamma_{\text{LM3}}(\epsilon)$.

\textbf{Step 2}: Constructing the sets $E_{j}$.

For each $j$, let 
\[
E_{j}:=C\cap B_{1}^{\text{on}}(\pi^{*},\frac{\epsilon_{\text{off}}\wedge\epsilon_{\text{LM3}}}{j})\cap B_{2}^{\text{on}}(\pi^{*},\frac{\epsilon_{\text{off}}\wedge\epsilon_{\text{LM3}}}{j})\cap B_{2}^{\text{off}}(\pi^{*},\epsilon_{\text{off}}).
\]
That is, $E_{j}$ is the set of strategy profiles that respect rational
compatibility, differ by no more than $\epsilon_{\text{off}}/j$ from
$\pi^{*}$ on path, and differ by no more than $\epsilon_{\text{off}}$
from $\pi^{*}$ off path. It is clear that each $E_{j}$ is convex
and compact, and that $\lim_{j\to\infty}E_{j}\subseteq B_{1}^{\text{on}}(\pi^{*},0)\cap B_{2}^{\text{on}}(\pi^{*},0)$
as claimed.

We may find an accompanying sequence of survival probabilities satisfying
\[
\gamma_{j}>\gamma_{\text{LM1}}(\delta,\frac{\epsilon_{\text{off}}\wedge\epsilon_{\text{LM3}}}{j})\vee\underline{\gamma}_{\text{LM2}}\vee\gamma_{\text{LM3}}(\frac{\epsilon_{\text{off}}\wedge\epsilon_{\text{LM3}}}{j})
\]
with $\gamma_{j}\uparrow1$.

\textbf{Step 3}: $\mathscr{R}^{g,\delta,\gamma_{j}}$ maps $E_{j}$
into itself.

Let some $\pi\in E_{j}$ be given.

By Lemma \ref{lem:sender_compatible} , $\mathscr{R}_{1}^{g,\delta,\gamma_{j}}[\pi_{2}]\in C$.

By Lemma \ref{lem:receiver1}, $\mathscr{R}_{2}^{g,\delta,\gamma_{j}}[\pi_{1}]\in B_{2}^{\text{off}}(\pi^{*},\epsilon_{\text{off}})$,
because uniformity of $\pi^{*}$ means $\text{BR}(\hat{P}(s),s)\subseteq\tilde{A}(s)$
for each off-path $s$.

By Lemma \ref{lem:receiver2}, $\mathscr{R}_{2}^{g,\delta,\gamma_{j}}[\pi_{1}]\in B_{2}^{\text{on}}(\pi^{*},\frac{\epsilon_{\text{off}}\wedge\epsilon_{\text{LM3}}}{j})$.

Finally, from Lemma \ref{lem:sender1} and the fact that $\pi_{2}\in B_{2}^{\text{on}}(\pi^{*},\frac{\epsilon_{\text{off}}\wedge\epsilon_{\text{LM3}}}{j})\cap B_{2}^{\text{off}}(\pi^{*},\epsilon_{\text{off}}),$
we have $\mathscr{R}_{1}^{g,\delta,\gamma_{j}}[\pi_{2}]\in B_{1}^{\text{on}}(\pi^{*},\frac{\epsilon_{\text{off}}\wedge\epsilon_{\text{LM3}}}{j})$.
\end{proof}

\end{document}